# Uniform taxation of electricity: incentives for flexibility and cost redistribution among household categories

Philipp Andreas Gunkel,[1*] Febin Kachirayil,[2] Claire-Marie Bergaentzlé,[1] Russell McKenna,[2,3] Dogan Keles[1] and Henrik Klinge Jacobsen[1]

1     Section for Energy Economics and Modelling, DTU Management, Technical University of Denmark, 2800 Kongens Lyngby, Denmark

2     Chair of Energy Systems Analysis, Institute of Energy and Process Engineering, ETH Zürich, 8092 Zürich, Switzerland

3     Paul Scherrer Institute, Laboratory for Energy Systems Analysis, Forschungsstrasse 111, 5232 Villigen PSI, Switzerland

*Correspondence: phgu@dtu.dk; Produktionstorvet 424, 2800 Kongens Lyngby, Denmark

**Abstract**. Recent years have shown a rapid adoption of residential solar PV with increased self-consumption and self-sufficiency levels in Europe. A major driver for their economic viability is the electricity tax exemption for the consumption of self-produced electricity. This leads to large residential PV capacities and partially overburdened distribution grids. Furthermore, the tax exemption that benefits wealthy households that can afford capital-intense investments in solar panels in particular has sparked discussions about energy equity and the appropriate taxation level for self-consumption. This study investigates the implementation of uniform electricity taxes on all consumption, irrespective of the origin of the production, by means of a case study of 155,000 hypothetical Danish prosumers. The results show that the new taxation policy redistributes costs progressively across household sizes. As more consumption is taxed, the tax level can be reduced by 38%, leading to 61% of all households seeing net savings of up to 23% off their yearly tax bill. High-occupancy houses save an average of €116 /year at the expense of single households living in large dwellings who pay €55 /year more. Implementing a uniform electricity tax in combination with a reduced overall tax level can (a) maintain overall tax revenues and (b) increase the interaction of batteries with the grid at the expense of behind-the-meter operations. In the end, the implicit cross-subsidy is removed by taxing self-consumption uniformly, leading to a cost redistribution supporting occupant-dense households and encouraging the flexible behavior of prosumers. This policy measure improves economic efficiency and greater use of technology with positive system-wide impacts.

**Keywords:** electricity tax; household characteristics; electrification; residential electricity consumption

## Nomenclature

### Abbreviations

    PV           Photovoltaic

| | |
|---|---|
| EV | Electric vehicle |
| HP | Heat pump |
| HST | Heat storage |
| BT | Battery |
| MP | Meter point |
| ToU | Time-of-Use |
| FIT | Feed-in tariff |
| RTP | Real-time pricing |
| VAT | Value added tax |
| BAU | Business as usual (scenario) |
| NTAX | New tax (scenario) |
| NTAX38 | New tax with 38% tax reduction (scenario) |
| NTAXHi | New tax with scaled electricity prices to 2022 (scenario) |

**Sets**

| | |
|---|---|
| H | Households |
| T | Hours |

**Variables**

| | | |
|---|---|---|
| $Q_{h,t}^{MP,imp,tot}$ | $[kWh]$ | Total grid import |
| $Q_{h,t}^{PV,tot}$ | $[kWh]$ | Total PV electricity production |
| $Q_{h,t}^{BT,MP}$ | $[kWh]$ | Flow from battery to meter point (export) |
| $Q_{h,t}^{PV,MP}$ | $[kWh]$ | Flow from PV to meter point (export) |
| $Q_{h,t}^{MP,CO}$ | $[kWh]$ | Flow from meter point to basic electricity consumption |
| $Q_{h,t}^{PV,CO}$ | $[kWh]$ | Flow from PV to basic electricity consumption |
| $Q_{h,t}^{BT,CO}$ | $[kWh]$ | Flow from battery to basic electricity consumption |
| $Q_{h,t}^{MP,BT}$ | $[kWh]$ | Flow from meter point to battery |
| $Q_{h,t}^{MP,EV}$ | $[kWh]$ | Flow from meter point to electric vehicle |
| $Q_{h,t}^{MP,HP}$ | $[kWh]$ | Flow from meter point to heat pump |
| $Q_{h,t}^{PV,BT}$ | $[kWh]$ | Flow from PV to battery |
| $Q_{h,t}^{PV,EV}$ | $[kWh]$ | Flow from PV to electric vehicle |
| $Q_{h,t}^{PV,HP}$ | $[kWh]$ | Flow from PV to heat pump |
| $SOC_{h,t}^{BT}$ | $[kWh]$ | State of charge of the battery |
| $Q_{h,t}^{BT,EV}$ | $[kWh]$ | Flow from battery to electric vehicle |
| $Q_{h,t}^{BT,HP}$ | $[kWh]$ | Flow from battery to heat pump |
| $SOC_{h,t}^{EV}$ | $[kWh]$ | State of charge of the electric vehicle |
| $Q_{h,t}^{HP,HST,He}$ | $[kWh^{Heat}]$ | Heat flow from heat pump to heat storage |
| $Q_{h,t}^{HP,CO,He}$ | $[kWh^{Heat}]$ | Heat flow from heat pump to heat consumption |
| $Q_{h,t}^{HST,CO,He}$ | $[kWh^{Heat}]$ | Heat flow from heat storage to heat consumption |
| $SOC_{h,t}^{HST,He}$ | $[kWh^{Heat}]$ | State of charge of the heat storage |

**Inputs/parameters**

| | | |
|---|---|---|
| $p_t^{el}$ | $[\frac{€}{kWh}]$ | Electricity price |
| $\tau^{tax}$ | $[\frac{€}{kWh}]$ | Electricity tax |
| $\tau^{grid}$ | $[\frac{\euro}{kWh}]$ | Volumetric grid tariff |
| $\beta^{N}$ | $[-]$ | Scenario binary |
| $q_{h,t}^{CO}$ | $[kWh]$ | Basic electricity demand |
| $cap^{MP}$ | $[kW]$ | Maximum capacity of meter point |
| $\eta^{CH,BT}$ | $[-]$ | Charger efficiency of the battery |
| $cap_{h}^{BT}$ | $[kWh]$ | Maximum storage capacity of the battery |
| $cap_{h}^{BT,CH}$ | $[kW]$ | Maximum charging/discharging capacity of the battery |
| $\beta_{h,t}^{EV}$ | $[-]$ | Availability variable of the electric vehicle |
| $\eta^{CH,EV}$ | $[-]$ | Charger efficiency of the electric vehicle |
| $v_{h,t}^{EV}$ | $[kWh]$ | Trip consumption of the electric vehicle |
| $cap^{EV}$ | $[kWh]$ | Maximum storage capacity of the electric vehicle |
| $cap^{EV,CH}$ | $[kW]$ | Maximum charging capacity of the electric vehicle |
| $\eta_{t}^{COP}$ | $[-]$ | COP of the heat pump |
| $\eta^{HST}$ | $[-]$ | Charger efficiency of the heat pump |
| $q_{h,t}^{CO,He}$ | $[kWh^{Heat}]$ | Heat demand |

## 1. Introduction

According to the policy goals of the Green New Deal published by the European Commission (European Commission, 2021, 2019), European residential consumers are expected to electrify their heating and transport demand in the coming years. Adopting heat pumps, electric vehicles, batteries, and solar PV transforms previously passive users of the electricity system into active prosumers. The increase in electricity consumption requires the installation of the new generating capacities previously found in central power plants, which must be added relatively soon to keep up with the adoption rates. At the same time, households are already beginning to install decentralized rooftop solar PV to reduce their net imports and cover their demand (Energistyrelsen, 2021). Governments are supporting the uptake of residential solar PV with several policies to increase local generation and greenhouse gas reductions to reach the climate goals (BDEW Bundesverband der Energie- und Wasserwirtschaft e.V., 2022).

The current policy environment in many European countries incentivizes residential solar PV deployment. Feed-in premiums for exported electricity improve the economic feasibility of residential solar PV (Arnold et al., 2022a; Fett et al., 2019). Behind-the-meter consumption of locally produced electricity for residential purposes is also exempt from the payment of electricity taxes and grid tariffs (BDEW Bundesverband der Energie- und Wasserwirtschaft e.V., 2022; Le gouvernement du Grand-duché de Luxembourg, 2021; Löfven, 2021; O'Donnell, 2019). This exemption drastically improves self-consumption's profitability compared to grid imports from the system while incentivizing behind-the-meter operations (Boampong and Brown, 2020; Fett et al., 2019). The current incentives encourage using individual flexibility to maximize the self-consumption of residential production to avoid tax and grid tariff payments that can amount to two thirds of the total residential electricity bill for grid imports (Eurostat, 2022). The focus on residential solar PV systems in combination with other flexible appliances that allow for behind-the-meter operation acts as a barrier to introducing flexibility into the system (Borenstein, 2022). In fact, the existing literature shows that the over-dimensioning of residential solar PV via favorable policies displaces flexible generation units and possibly leads to increased system costs and lower reliability of the electricity system (Simshauser, 2022). The tax and grid tariff exemptions for residential solar PV reduces the income for system-relevant services, the resulting shortfall having to be recovered from the remaining grid consumers (Borenstein, 2017). Households that are dependent on grid electricity often lack the equity to invest in residential solar PV and face higher charges from system operators and states to recover the losses from taxes and grid tariffs, which can lead to marginalization and a downward spiral (Borenstein et al., 2021; Crago et al., 2023; European Union, 2022).

While studies already point out that certain levels of adoption due to subsidies impose problems for the electricity grid and system and lead to social disparities, the impact of the tax exemption on self-consumption from domestically produced electricity is only mentioned theoretically and is not quantified. Therefore, this study investigates the impact of uniform electricity taxes on all consumption, including behind-the-meter self-consumption. From the overall objective follow the following two sub-research questions:

- How are the taxes and total electricity costs redistributed among household categories?
- What is the impact on the flexible behavior of prosumers in terms of self-consumption and grid interaction and, subsequently, system efficiency?

This study uses a linear optimization framework to optimize the operation of flexible technologies in prosumer households using heat pumps, heat storage, solar PV systems, batteries, and electric vehicles individually. The case study covers in unprecedented detail 155,000 Danish households divided into 36 socio-economic categories with different electricity, heat and EV consumption patterns. The implementation of a distributed and parallelized code offers reduced computational time while providing robust results for every relevant prosumer category in Denmark. This study contributes to the literature by:

- Developing of a scalable methodology with detailed household data to reflect the heterogeneity of individual households at the national level
- Breaking down the impact of uniform electricity taxes on households depending on their socio-economic characteristics
- Delivering perspectives for future policy discussions on the taxation of consumers and prosumers

The study is structured as follows. Section 2 provides a literature review focusing on modeling prosumer households in the context of changing policy designs. Section 3 introduces the taxation methodologies, state-of-the-art optimization frameworks and the mathematical formulation of the optimization model used. The fourth section presents the case study's data for Denmark, followed by a definition of the studied scenarios. Section 5 summarizes the distribution of electricity costs across all households and further disaggregates these results in line with socio-economic categories. It also explores the impact on power flows and the sensitivity of these results to the high electricity prices experienced in 2022. Section 6 discusses the outcomes and puts them into perspective by comparing them with the literature. Then the outcomes are generalized into a European context, and policy recommendations are formulated after that. Section 7 concludes the study and outlines future work.

*2 Literature review*

This literature review focuses first on policies such as subsidies, grid tariffs, and taxes and their impact on different actors. Second, the review highlights modeling approaches for prosumer households focusing on the representation and effects of policy environments. This study investigates the effect of redesigning electricity tax rate differentials within households, while the literature mainly focuses on aggregate tax levels, subsidies, grid tariffs, and different metering designs. At the same time, integrating grid tariff policies in prosumer models is similar to current electricity tax designs in mainly being volumetric and applied to net consumption. Subsequently, the literature review mainly focuses on those studies and their impact on cost, flexibility, and PV integration.

*2.1 Reviewing the impact of policies on prosumer and solar PV adoption*

One of the main approaches used in Europe and other countries has been to improve solar PV's economic feasibility via subsidies, favorable grid tariffs, and tax designs. Germany has implemented a subsidy system guaranteeing fixed feed-in tariffs for residential solar PV systems (Dietrich and Weber, 2018). Solar PV systems with a size of up to 30 kWp received the maximum rates for electricity exports. At the same time, self-consumption is not subject to taxes, grid tariffs or other levies (BDEW Bundesverband der Energie- und Wasserwirtschaft e.V., 2022). Trends to increase the sizes of solar PV thresholds and the tax exemptions for self-consumed electricity are seen across Europe, e.g. France (O'Donnell, 2019), Sweden (Löfven, 2021) and Luxemburg (Le gouvernement du grand-duché de Luxembourg, 2021). Denmark, however, considered a proposal in parliament in 2017 to reduce the maximum size of solar PV that receives support and the subsequent tax exemption (Energi- forsynings og klimaministerien, 2017). The faster-than-expected uptake of residential solar PV reduced the estimated revenue from electricity taxes, which are used to fund basic state services, by €160 M (Emiliano Bellini, 2017). Arnold et al. show that high feed-in tariffs, in combination with the rising value of self-consumption driven by taxes, levies and grid tariff designs, increase the adoption of residential solar PV (Arnold et al., 2022b). Moreover, Gautier et al. find that, e.g., net metering in support of PV adoption leads to even bigger economic distortions to the benefit of prosumers but at the expense of consumers (Gautier et al., 2018). The outcomes further show a high correlation between household income and avoided taxes and indicate the benefits from disconnecting from the system. Crago et al. even show that different ownership models of solar PV, depending on the financial status of the household, increase the divide between vulnerable and well-off households in the US (Crago et al., 2023). Thus, subsidies for residential solar systems are increasingly being discussed as controversial, e.g. by Borenstein (Borenstein, 2022), and (Grösche and Schröder, 2014). The first author argues that focusing on behind-the-meter applications of self-produced solar PV did not provide the grid-facing flexibility from the consumer side that was hoped for in California.

The adoption of residential solar PV supported by policy incentives has considerable effects on the overall energy system and markets. One such case is being investigated in Queensland, Australia, by Simshauser et al. (Simshauser, 2022). In sum, the authors' analysis shows that the rapid adoption of residential PV displaces flexible generation units, while PV systems usually do not deliver the required production during hours of peak demand. The implementation of feed-in tariffs, higher network tariffs, encouragement of self-consumption, and installation of overcapacity through investment incentives can cause electricity prices to drop to negative levels, as is discussed by Simshauser (Simshauser, 2022). Regions with higher shares of rooftop solar encounter challenges to limited distribution grid capacities from reduced network revenue contributions (Borenstein, 2022, 2017) as well as from rapid changes in the dynamics of established markets dealing with shortages of flexible capacity due to displacement (Simshauser, 2022). To give appropriate economic signals, (Yu, 2021) suggests an enhanced market design that includes additional payment systems with high shares of residential solar PV. While grid tariffs, subsidies, market frameworks and flexibility are investigated in the literature, the role of distortions in electricity taxes represents a gap in analyzing the optimal integration of solar PV on the social level.

Electricity tariffs recover costs for system services such as reliability, transmission, and distribution, whereas electricity tax revenues are fiscal to fund basic state services. The common approach to network cost recovery using volumetric surcharges on electricity renders the system regressive, as electricity consumption does not scale linearly with income (Mastropietro, 2019), while self-consumption makes it possible to avoid contributing to system services (Fett et al., 2019). On the other hand, reducing electricity tax levels would help smaller and poorer households, as Borenstein demonstrated for California (Borenstein et al., 2021). While these studies focus mainly on the United States or more on grid tariffs, the additional role of varying electricity tax rates across consumer categories is not considered. Fausto et al., however, outline the distorting effects of such uneven tax and grid tariff designs (Fausto et al., 2019). They see the current design of electricity taxes and tariffs as unsuitable for the future mix of consumers and prosumers.

*2.2 Prosumer modelling approaches, including policy designs*

An economic assessment is usually at the center of studies modeling prosumer households, from the perspective of either the prosumer (Castellini et al., 2021; Fett et al., 2019; Pena-Bello et al., 2017) or the energy system (Boccard and Gautier, 2021; Freitas Gomes et al., 2021; Gautier et al., 2018). The significance of the remuneration scheme and the savings from self-consumption for the viability of prosuming has been demonstrated both in models (Dietrich and Weber, 2018) and empirically (Arnold et al., 2022b). Therefore, different grid-tariff designs are discussed, including in how they influence the incentives of prosumers.

Different tariff structures are investigated in the literature, including fixed rates that do not tak actual household electricity consumption into account, volumetric tariffs (€/kWh) and peak-demand charges (€/kW), as well as combinations of these three options (Simshauser, 2016). Volumetric charges can be flat if the rate is constant over time, reflect time-of-use (ToU) if there is a lower off-peak rate and a higher on-peak rate, or reflect real-time pricing (RTP) if rates vary by hours (Ansarin et al., 2020; Pena-Bello et al., 2017). Demand charges require smart meters to identify the peak demand (Manuel de Villena et al., 2021; Simshauser, 2016). Furthermore, the time horizon over which the peak demand is determined has to be chosen. Options include the consumer's private maximum consumption, the maximum private consumption during hours when the system is constrained or the private maximum during a given set of hours akin to ToU (Boampong and Brown, 2020). Volumetric tariffs are the most common way to recover costs for promoting renewable energy sources (Mastropietro, 2019). Overall, different designs alter optimal operation and investments, while most studies focus only on the optimization from the private view and do not include the system.

(Heinisch et al., 2019) demonstrate the importance of aligning prosumer incentives with system incentives by optimizing the same PV and battery system from each actor's perspective. From a system perspective, the optimal operation uses household-level batteries to avoid peak-unit generation and the curtailment of non-dispatchable generation. From a household perspective, on the other hand, the optimal battery-use case is to match the solar generation profile to the load profile. That allows the prosumer to avoid electricity taxes and grid fees, which is the objective of this study. (Boampong and Brown, 2020) find that adding a demand charge to a volumetric tariff based on the household's peak demand during hours when the system is constrained can avoid this issue and align household incentives with the system's interests.

A switch from the commonly used net-metering approach to net-purchasing can have similar benefits by encouraging self-consumption and investment in more battery capacity (Manuel de Villena et al., 2021). Indeed, net-metering was found to produce too many prosumers (Gautier et al., 2018). This shift can also increase social welfare by reducing cross-subsidies from consumers to prosumers (Picciariello et al., 2015).

While the emphasis of the previous studies was on the operation of exogenously sized PV and battery systems (Fett et al., 2019), optimize the PV and battery capacities from a household perspective under different changes to the regulatory framework. They find that a feed-in tariff is the most effective tool to encourage PV generation. However, they also show that the cost recovery for the FIT leads to the highest increase in electricity prices because support is financed differently in the two cases. As (Gautier et al., 2018) found, this leads to a decrease in prosumer expenses at the expense of consumers.

Volumetric surcharges lead to cross-subsidies for prosumers at the expense of consumers, which are commonly used to recover costs in the energy system (Burger et al., 2020; Mastropietro, 2019). (Burger et al., 2020) propose shifting to progressively fixed charges to avoid cross-subsidies, stating that, while fixed charges lead to disparities as poorer households consume less electricity, the cost recovery through volumetric charges leads to an even bigger economic distortion. Adding peak demand charges to a two-part tariff structure that already includes fixed and volumetric charges has also been found to reduce cross-subsidies (Simshauser, 2016; Strielkowski et al., 2017). (Ansarin et al., 2020) compare five different tariffs – a flat volumetric charge, flat with tiered volumetric prices, ToU pricing, hourly RTP and hourly RTP with a demand charge – to assess the cross-subsidies within each scenario. They find that RTP performs best with near-zero cross-subsidies, while a ToU tariff yields approximately USD 75 in cross-subsidies, and the combination of RTP with a monthly demand charge costs the worst-impacted household USD 900 more than the expenditure it is responsible for. However, all of these pale in comparison with the two flat volumetric rates, which result in USD 3000 and USD 4000 in cross-subsidies for the worst-impacted households respectively, as they redistribute the fixed charges on a volumetric basis. From the opposite angle (Pena-Bello et al., 2017), find that a flat tariff is more beneficial to prosumers than ToU or RTP. Indeed, further cross-subsidies and the reason for these socio-economic disparities is that wealthier households tend to be more frequent house owners, own bigger homes, and consume more electricity, all of which make the investment more attractive (De Groote et al., 2016).

(Tomasi, 2022) studies the effect of the policy initiative "The tax on sun" in Spain that introduced several barriers to the residential adoption of PV, including a tax on self-consumed electricity from domestic solar PV. Using a synthetic analysis, the authors compared the current electricity production of PV with a hypothetical production without the additional tax burden. They conclude that the "tax on the sun" leads to a small reduction in total PV production but acts as a barrier to the higher adoption of residential solar PV systems.

The literature review provided the background of policies improving the business case for residential PV, as well as the challenges that come with it. While several policies were investigated, the role of electricity taxation, in particular in Europe, was neither covered nor modelled. Following the literature review on prosumer household modeling, the following section introduces the applied methodology, including the electricity tax design and the mathematical modeling approach.

*3. Methodology*

The methodology section first introduces the current tax design and summarizes the adapted tax design based on (Fausto et al., 2019). The last subsection introduces the mathematical formulation of the used optimization model. Figure 1 provides a schematic overview of the data and model structure.

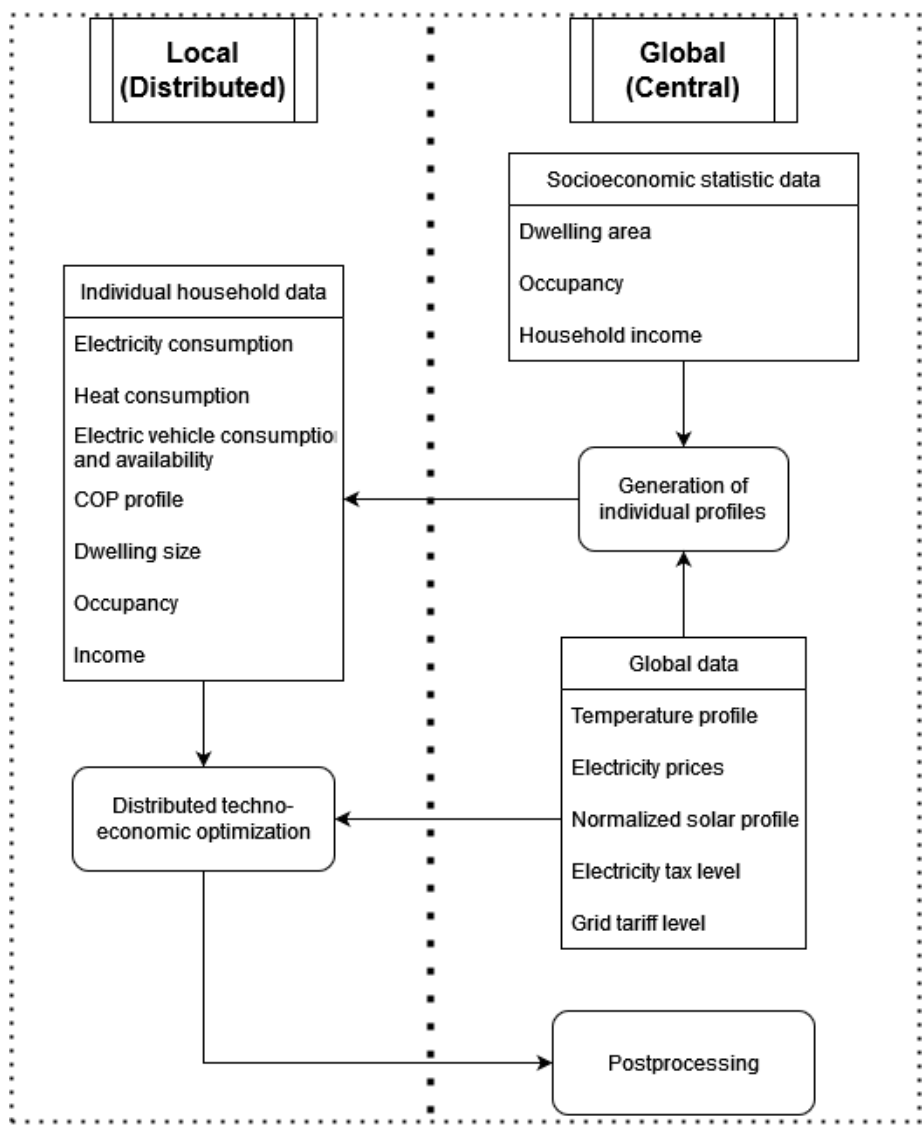

*Figure 1. Schematic overview of the interplay of data input and the optimization model used in this study.*

*3.1 Electricity taxation in Europe, using Denmark as an example of the relevance of taxes in the optimization analysis*

This study follows the taxation scheme in Denmark to illustrate the importance of varying tax rates and network tariffs,. Europe has several taxes, levies and surcharges on net withdrawals. The map in Figure 2 shows the aggregated charges on net withdrawal, excluding electricity, supply and network costs, in ct/kWh.

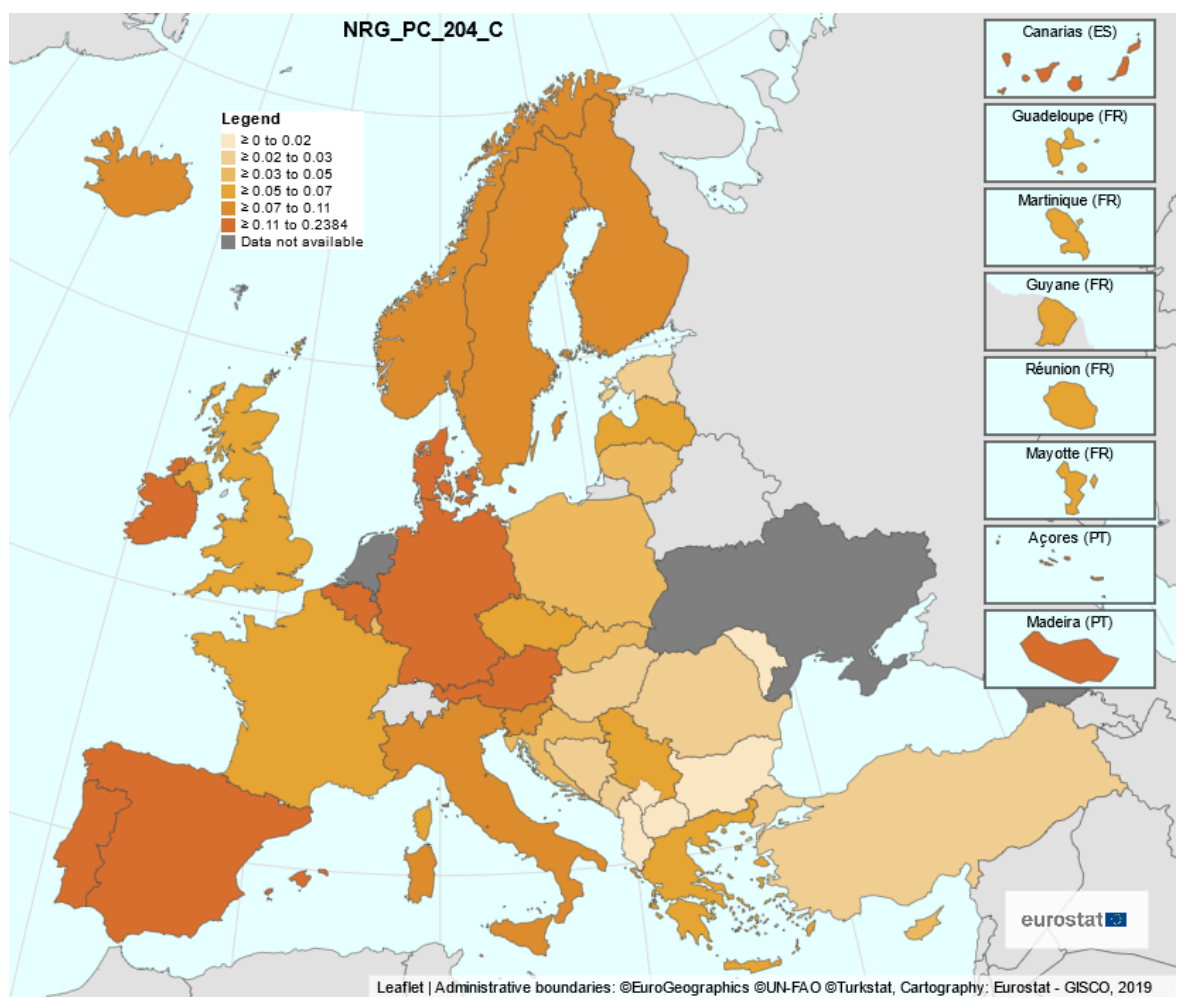

*Figure 2. Average electricity taxation, fees and subsidies in ct/kWh in Europe, excluding the cost of electricity supply and network charges* (Eurostat (Online data: nrg_d_hhq), 2022).

The map shows that especially countries such as Denmark, Germany, Belgium, Spain, Portugal and Austria have high additional charges for electricity consumption from the grid. Most countries have electricity taxes, fees and subsidies applied as volumetric charges. On the household level, they usually apply to grid consumption, not self-consumption from residential generation via PV. As soon as the installed PV passes a certain country-specific threshold or the household decides to draw a tax credit on the VAT, households are subject to the named taxes, fees and subsidies on self-consumption. The EU Directive from 2018 specifically mentions the exclusion of electricity charges and taxes to promote self-consumption and accelerate the transition in Europe (European Commission, 2018). At the same time, the directive acknowledges that this should not be done by all means, proposing instead a maximum threshold of 30kWp and further allowing taxes to be levied to support the "financial sustainability of the energy system".

Irrespective of the maximum thresholds, which differ across countries, electricity taxes apply very similarly to households that own a solar PV below the threshold. Figure 3 below visualizes all electricity and energy flows (arrows) in combination with the cost components connected to those (colors) where taxes, electricity costs and grid tariffs apply.

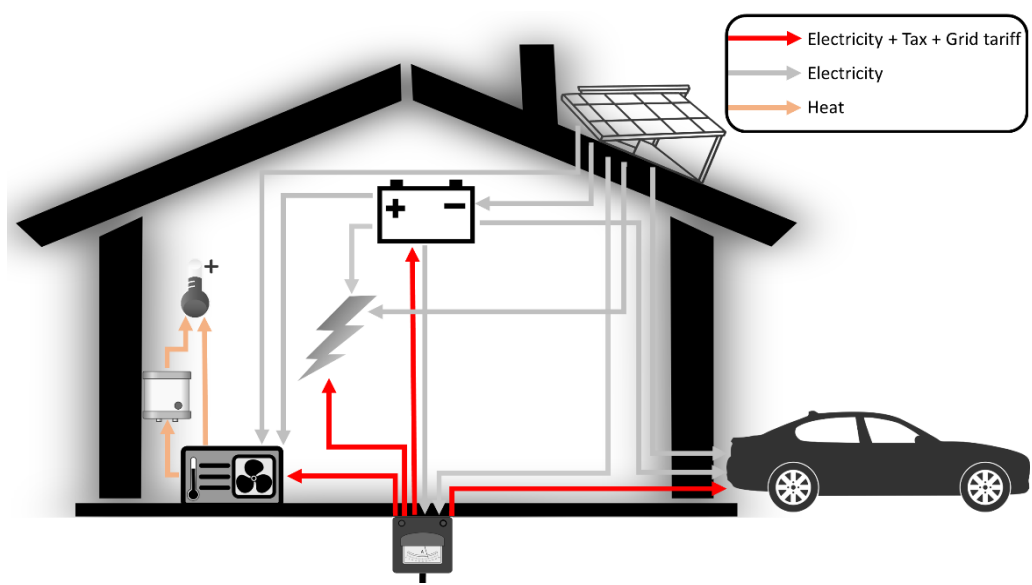

*Figure 3. Energy flows in households and currently applied policies in our study. The grey lines represent electricity flows that are neither taxed nor subject to grid tariffs, the red lines show flows subject to both taxes and grid tariffs. The orange arrows represent heat flows.*

All flows from the meter point in the bottom going to the heat pump, covering basic consumption, including charging the battery or the electric vehicle, are subject to electricity costs, taxes and grid tariffs. Conversely, electricity flows from rooftop solar PV are only connected to the marginal cost of the technology. Providing flexibility through the battery by offering arbitrage is disincentivized by double taxation, as taxes are paid when the household does arbitrage trading when importing power and then again by the end consumer when he consumes the electricity exported from the battery. In the remainder of this study, Business As Usual (BAU) refers to this current tax scheme applied in most (European) countries.

Figure 4 visualizes the new tax scheme according to the work of (Fausto et al., 2019).

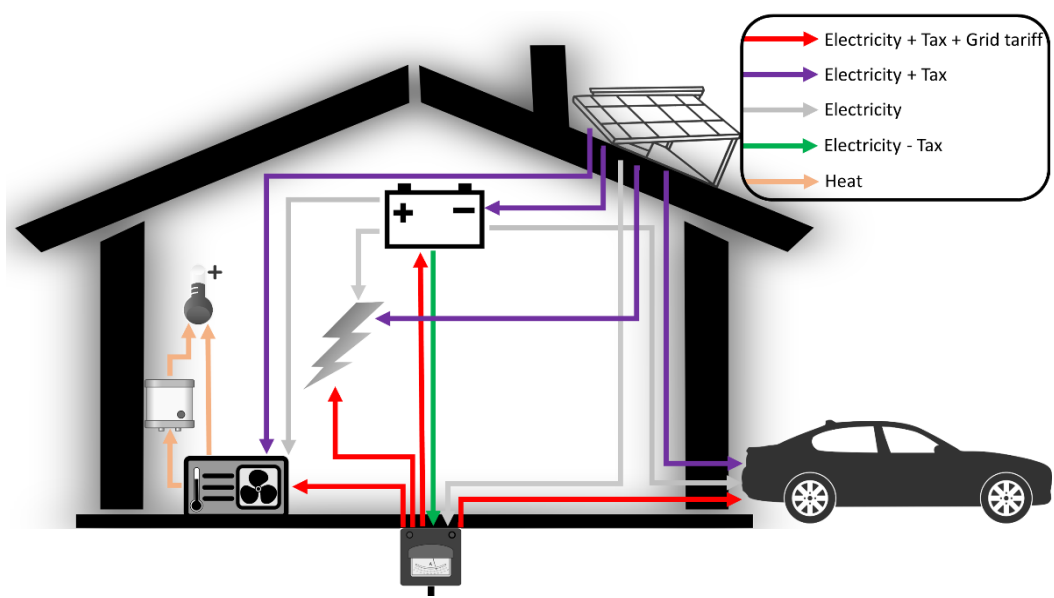

*Figure 4. Energy flows in households in the NTAX policy scenario. The grey lines represent electricity flows that are neither taxed nor subject to grid tariffs, the red lines shows flows subject to both taxes and grid tariffs, the green lines are subject to tax refunds, and the electricity price and purple lines are taxed. For the sales from the battery to the grid (green arrow), the previously paid taxes are refunded. The orange arrow represents heat flows.*

The focus of the electricity tax is the non-distortionary collection of tax revenue for public funding. This implies that the tax should be broadly applied, with an equal rate on all types of own consumption. Grid imports are still subject to both grid tariffs and taxes. Flows from the PV to technologies or demand inside the house are now subject to taxes, whether they are used immediately or stored in the battery. Direct exports of domestic production remain free of taxes to avoid double taxation. This principle is also used for exports from the battery, which lead to a refund of previously paid taxes. The tax refund applies only to the quantity of energy exported from the battery (rate per kWh), irrespective of whether or not the value of exported electricity is higher than imported. The energy flow is subject to the full taxes for the battery charging from either domestic production from PV or imports. Only flows where a tax was paid qualify for a tax refund, so grid imports qualify, but not direct exports from self-generation. The taxation scheme thus incentivizes efficient technologies and electricity. In the remainder of this study, the taxation scheme visualized in Figure 4 is called NTAX (referring to new tax).

*3.2 Mathematical formulation of the optimization model*

We formulate an optimization model focusing on the operation of the equipment under least cost while fulfilling all technical constraints. The first subsection presents the objective function, including all cost elements. The second subsection covers all technical constraints, including household energy flows and balances. Exogenous inputs are explained and presented in Section 4.1.

*3.2.1 Objective function*

The households' objective function covers all operational costs associated with meeting basic electricity, heat, and EV demand as outlined in Equation (1) for the BAU and NTAX scenarios.

$$\min \sum_{t}^{T} p_t^{el} Q_{h,t}^{MP,imp,tot} + \tau^{grid} Q_{h,t}^{MP,imp,tot} + \tau^{tax}(Q_{h,t}^{MP,imp,tot} + \beta^N Q_{h,t}^{PV,tot}) - p_t^{el}(Q_{h,t}^{BT,MP} + Q_{h,t}^{PV,MP}) \quad (1)$$

$$- \beta^N \tau^{tax}(Q_{h,t}^{BT,MP} + Q_{h,t}^{PV,MP}) \qquad \forall \ h \ \epsilon$$

The total cost for each household $H$ and hourly timesteps $T$ is minimized. $p_t^{el}$ represents the hourly varying electricity price. $\tau^{grid}$ and $\tau^{tax}$ are constant volumetric grid tariffs and electricity taxes applied on electricity imports $Q_{h,t}^{MP,imp,tot}$ as measured by the meter point. Revenue is generated by "exporting" electricity (selling to the grid) from the battery $Q_{h,t}^{BT,MP}$ and from the PV $Q_{h,t}^{PV,MP}$. The binary $\beta^N$ is zero for the BAU scenario.

The objective function of the NTAX scenario includes additional terms compared to the BAU scenario, indicated by the binary $\beta^N$ that switches to one. In addition to imports, all production of solar PV $Q_{h,t}^{PV,tot}$ is also subject to grid tariffs and electricity taxes. At the same time, a tax refund applies to electricity exported from the battery $Q_{h,t}^{BT,MP}$ or from solar PV $Q_{h,t}^{PV,MP}$. The tax level $\tau^{tax}$ stays the same for NTAX. An additional scenario, NTAX38, has a reduced tax level $\tau^{tax}$.

*3.2.2 Constraints*

Equations (2) and (3) describe the supply of basic electricity demand $q_{h,t}^{CO}$ in the household.

$$q_{h,t}^{CO} = Q_{h,t}^{MP,CO} + Q_{h,t}^{PV,CO} + Q_{h,t}^{BT,CO} \qquad (2)$$

$$Q_{h,t}^{MP,CO}, Q_{h,t}^{PV,CO}, Q_{h,t}^{BT,CO} \geq 0 \qquad (3)$$

With $Q_{h,t}^{MP,CO}$ being flows from the meter point, $Q_{h,t}^{PV,CO}$ representing the supply from PV production and $Q_{h,t}^{BT,CO}$ flows from the battery.

Total electricity imports from the grid are a sum of the specific flows from the meter point to the technologies. This sum $Q_{h,t}^{MP,imp,tot}$ is calculated in Equations (4)-(6).

$$Q_{h,t}^{MP,imp,tot} = Q_{h,t}^{MP,BT} + Q_{h,t}^{MP,EV} + Q_{h,t}^{MP,HP} + Q_{h,t}^{MP,CO} \qquad (4)$$

$$cap^{MP} \geq Q_{h,t}^{MP,imp,tot} \qquad (5)$$

$$Q_{h,t}^{MP,BT}, Q_{h,t}^{MP,EV}, Q_{h,t}^{MP,HP} \geq 0 \qquad (6)$$

While $Q_{h,t}^{MP,BT}$ is the imports of electricity from the grid needed to charge the battery, $Q_{h,t}^{MP,BT}$ charges the EV directly from the grid, $Q_{h,t}^{MP,HP}$ sends the electricity to the HP, and $Q_{h,t}^{MP,CO}$ covers basic electricity demand. The main fuse in houses determines the maximum import $cap^{MP}$, which is around 21-28 kW in Denmark (Radius El-net, 2022).

Total PV production corresponds to the sum of all flows to household technologies or to export $Q_{h,t}^{PV,MP}$ through the meter point, as shown in Equations (7)-(9).

$$Q_{h,t}^{PV,tot} \eta^{PV} = Q_{h,t}^{PV,BT} + Q_{h,t}^{PV,EV} + Q_{h,t}^{PV,HP} + Q_{h,t}^{PV,CO} + Q_{h,t}^{PV,MP} \qquad (7)$$

$$cap_h^{PV} \geq Q_{h,t}^{PV,tot} \geq 0 \qquad (8)$$

$$Q_{h,t}^{PV,BT}, Q_{h,t}^{PV,EV}, Q_{h,t}^{PV,HP}, Q_{h,t}^{PV,MP} \geq 0 \qquad (9)$$

With $Q_{h,t}^{PV,BT}$ and $Q_{h,t}^{PV,EV}$ standing for the PV production charging the battery and EV respectively, $Q_{h,t}^{PV,HP}$ showing the flow to the HP, and $Q_{h,t}^{PV,CO}$ covering the basic electricity demand of the house. The PV production depends on the normalized solar irradiation $\eta^{PV}$ ranging between zero and one. The installed capacity $cap_h^{PV}$ restricts the maximum PV production, which depends on the house size.

The constraints of the battery cover two main parts. First, the energy storage equations in Equations (10)-(11) describe the energy balance; second, Equations (12)-(13) cover the charging and discharging of the battery.

$$SOC_{h,t}^{BT} = SOC_{h,t-1}^{BT} + (Q_{h,t}^{MP,BT} + Q_{h,t}^{PV,BT})\eta^{CH} - (Q_{h,t}^{BT,CO} + Q_{h,t}^{BT,EV} + Q_{h,t}^{BT,HP} + Q_{h,t}^{BT,MP})\frac{1}{\eta^{CH}} \qquad (10)$$

$$cap_h^{BT} \geq SOC_{h,t}^{BT} \qquad (11)$$

$$cap_h^{BT,CH} \geq Q_{h,t}^{BT,CO} + Q_{h,t}^{BT,EV} + Q_{h,t}^{BT,HP} + Q_{h,t}^{BT,MP} + Q_{h,t}^{MP,BT} + Q_{h,t}^{PV,BT} \qquad (12)$$

$$Q_{h,t}^{BT,EV}, Q_{h,t}^{BT,HP}, Q_{h,t}^{BT,MP} \geq 0 \qquad (13)$$

$SOC_{h,t}^{BT}$ is the state of charge of the battery, which is restricted by the maximum exogenous building-specific capacity $cap_h^{BT}$. The charging from the meter point and PV, $Q_{h,t}^{MP,BT}$ and $Q_{h,t}^{PV,BT}$, and discharging to cover basic consumption $Q_{h,t}^{BT,CO}$, EV demand $Q_{h,t}^{BT,EV}$, the heat pump $Q_{h,t}^{BT,HP}$ and exports $Q_{h,t}^{BT,MP}$ are subject to efficiencies represented by $\eta^{CH}$. $cap_h^{BT,CH}$ limits the maximum charging and discharging capacity.

The EV has similar constraints to the battery. At the same time, the availability of the vehicle and, subsequently, the charging is restricted by trips. Equation (14)-(19) describe the energy balance of the EV and its charging.

$$SOC_{h,t}^{EV} \beta_{h,t}^{EV} = SOC_{h,t-1}^{EV} + (Q_{h,t}^{MP,EV} + Q_{h,t}^{PV,EV} + Q_{h,t}^{BT,EV})\eta^{CH,EV} - v_{h,t}^{EV} \qquad (14)$$

$$cap^{EV} \geq SOC_{h,t}^{EV} \qquad (15)$$

$$SOC_{h,t}^{EV} = cap_h^{EV} \quad \forall \ t \ \epsilon \ T_{Departure} \qquad (16)$$

$$cap^{EV,CH}\beta_{h,t}^{EV} \geq Q_{h,t}^{MP,EV} + Q_{h,t}^{PV,EV} + Q_{h,t}^{BT,EV} \tag{17}$$

$$Q_{h,t}^{MP,EV} + Q_{h,t}^{PV,EV} + Q_{h,t}^{BT,EV} \geq q_{h,t}^{EV,forced} \tag{18}$$

$$Q_{h,t}^{BT,EV}, SOC_{h,t}^{EV} \geq 0 \tag{19}$$

$SOC_{h,t}^{EV}$ is the state of charge of the battery, which is multiplied by the exogenous availability binary $\beta_{h,t}^{EV}$. The binary equals one when the vehicle is at home and zero when the vehicle is on a trip. Electricity flows for charging the EV come from the meter point, PV or battery. $\nu_{h,t}^{EV}$ represents the consumption of a trip. The vehicles' battery size $cap^{EV}$ determines the maximum state of charge of the EV. All EVs have to be fully charged one hour before departure, as determined by $T_{Departure}$. Furthermore, the installed charger capacity $cap^{EV,CH}$ restricts charging flows. At the same time, the vehicle availability binary also determines if charging is possible. EVs can arrive from a trip with a state of charge that is insufficient to satisfy emergency trips to e.g. hospitals. In those circumstances, an exogenous charging parameter $q_{h,t}^{EV,forced}$ forces the battery to be charged to meet the minimum state of charge requirements following the methodology of (Gunkel et al., 2020).

Equations (20)-(26) describe the house's heating system, including the heat pump and heat storage operation.

$$Q_{h,t}^{HP,HST,He} + Q_{h,t}^{HP,CO,He} = (Q_{h,t}^{MP,HP} + Q_{h,t}^{PV,HP} + Q_{h,t}^{BT,HP})\eta_t^{COP} \tag{20}$$

$$cap_h^{HP} \geq Q_{h,t}^{MP,HP} + Q_{h,t}^{PV,HP} + Q_{h,t}^{BT,HP} \tag{21}$$

$$SOC_{h,t}^{HST,He} = SOC_{h,t-1}^{HST,He} + Q_{h,t}^{HP,HST,He}\eta^{HST} - Q_{h,t}^{HST,CO,He}\frac{1}{\eta^{HST}} \tag{22}$$

$$cap_h^{HST} \geq SOC_{h,t}^{HST,He} \tag{23}$$

$$q_h^{HST,CH} \geq Q_{h,t}^{HP,HST,He}, Q_{h,t}^{HST,CO,He} \tag{24}$$

$$q_{h,t}^{CO,He} = Q_{h,t}^{HST,CO,He} + Q_{h,t}^{HP,CO,He} \tag{25}$$

$$Q_{h,t}^{HP,HST,He}, Q_{h,t}^{HP,CO,He}, Q_{h,t}^{HP,HST,He} \geq 0 \tag{26}$$

The heat production of the heat pump can either go to the heat storage $Q_{h,t}^{HP,HST,He}$ or cover the basic heat demand $Q_{h,t}^{HP,CO,He}$ and is restricted by the installed heat pump capacity $cap_h^{HP}$. Electricity supplying the heat pump is converted into thermal energy using the time-variant conversion factor $\eta_t^{COP}$. The heat storage state of charge $SOC_{h,t}^{HST,He}$ is restricted by the exogenously defined installed capacity $cap^{HST}$, and charging and discharging is subject to efficiencies $\eta^{HST}$. The charging and discharging capacity $q_h^{HST,CH}$ limits energy flows from and to the heat storage represented by $Q_{h,t}^{HP,HST,He}$ and $Q_{h,t}^{HST,CO,He}$ respectively. The combination of energy supply from the heat pump and the heat storage covers the basic heat demand $q_{h,t}^{CO,He}$.

*4. Case study*

Based on the Danish system's characteristics, this study investigates the impact of taxing self-consumed solar production in a sample scenario. Denmark has set ambitious targets for decarbonizing the energy it uses and is currently more or less on track to meet the requirements of the Paris agreement (Energistyrelsen, 2021). The Danish Energy Agency foresees rapid growth of solar PV of around 650% by 2030 compared to the 2020 level (Energistyrelsen, 2020). Most of this additional PV capacity should be located in areas with residential rooftops. The decarbonization of residential heating is another central policy objective of the Danish government, which has introduced a reduction of electricity tax levels in combination with a ban on natural gas boilers by 2035 in order to decarbonize the residential heating sector entirely. Similarly, the uptake of electric vehicles is being incentivized by imposing very low levels of registration duties on such vehicles.

The anticipated rapid adoption of HPs, EVs and PV in residential buildings changes the energy landscape and how electricity costs are distributed across various consumer categories. We assume that consumers aim to reduce their energy expenses and that large consumption technologies like the aforementioned heat pumps and electric vehicles can be controlled automatically. The following two subsections first present the underlying residential data of Danish prosumer households and assumptions. Second, the four scenarios that have been investigated and are analyzed are introduced.

### 4.1 Danish household data

This study categorizes residential households according to (F. M. Andersen et al., 2021; Gunkel et al., 2022) in order to model the impact of the new tax design. Both studies follow the "Manual for statistics on energy consumption in households" issued by Eurostat to provide international standards and comparability of results (European Union, 2013). Hourly residential electricity consumption measured by smart meters is linked to socio-economic consumer categories (F. M. Andersen et al., 2021; Frits Møller Andersen et al., 2021; Gunkel et al., 2022). The study uses electricity data from before the Covid-19 pandemic and the gas shortage in Europe, but also before real-time pricing and Time-of-Use tariffs were broadly implemented in Denmark. Table 1 summarizes the characteristics that describe the analyzed categories.

| Characteristic name | Characteristics | | | |
|---|---|---|---|---|
| Dwelling type | H: House | | | |
| Occupancy | P1: 1 occupant | P2: 2 occupants | P3: 3-4 occupants | P5+: 5 or more occupants |
| Dwelling area | A1<$110m^2$ | $110m^2$<A2<$146m^2$ | $146m^2$<A3 | |
| Income level | €1<$240kDKK$ | $240kDKK$<€2<$449kDKK$ | $449kDKK$<€3 | |

*Table 1 Chosen socio-economic categories and their respective values.*

This study focuses only on single-family houses and semi-detached houses that act as prosumers due to the appliances' clear connection/ownership and operational status and to avoid shared energy community policy frameworks, as well as those that differ by country. Occupancy rates are divided into singles (P1), two persons (P2), three to four persons (P3), and five or more persons (P5+). The dwelling area is divided into three categories using quantile statistics (terciles – a third of the population per group). A1 represents houses that have an area of up to 110m$^2$, (A2) between 110m$^2$ and 146m$^2$, and (A3) more than146m$^2$. The households are further divided by total household income using median statistics. The lower third of the Danish population earns up to DKK 240.000 (€32172) per year, indicated by (€1). The (€2) households earn between DKK 240.000 and DKK 449.000 (€60188) per year, whereas (€3) households earn more than DKK 449.000 per year. All characteristics are mixed to form a total of 35 categories, each connected to 1000 randomized synthetic profiles of electricity consumption. Each household is assumed to own solar PV and a battery, both of which are scaled depending on the dwelling area. The battery is scaled assuming 1kWh of storage per 1.375 kWp installed solar PV, and the PV size scales with 0.039 kWp per m$^2$ of the house following the results of previous studies (DiOrio et al., 2020; Weniger et al., 2014; Yang et al., 2018). The marginal cost of producing electricity with a residential solar panel is taken to be zero.

Each household has a dwelling area connected to it, which is used for the dwelling area categorization (A1-A3). Even though the heated area is the primary indicator of total heat consumption, the house's insulation plays a major role. We use actual heating data of (Siddique et al., 2022) covering Danish households and divide each dwelling area category (A1-A3) into three building ages that are used as a proxy to differentiate heat consumption per m$^2$ based on the insulation standards of each period. Figure 10 in the appendix summarizes all combinations of dwelling areas and building ages with their respective yearly heat demand in this study. Heat pumps and heat storage supply the heat for each building. The size of the heat pump and heat storage follows the guidelines of the Danish Energy Agency and the Danish Technological Institute, resulting in 0.031 kW per m$^2$ for heat pumps and 0.045 kWh of storage capacity per m$^2$ of hot water tanks (Poulsen et al., 2017).

Every household is further assumed to own an electric vehicle that uses smart charging, i.e. whose charging time is flexible. Here consumption varies between 1800 kWh and 3200 kWh per year, as shown in Figure 11 in the Appendix. The vehicles' efficiency depends on the daily temperature profile. The availability of each vehicle for home charging is shown in Figure 12 in the Appendix. The consumption and vehicle availability dataset is taken from (Gunkel et al., 2020) and is based on the Danish National Transport Survey (Christiansen and Warnecke, n.d.).

The study focuses on Denmark, which has relatively small spatial temperature differences due to its size and coastal climate conditions. The temperature curve is taken from the Danish Meteorological Institute for the city of Vejle for heat and EV demand (Cappelen, 2019; Danmarks Meteorologiske Institut (unofficial), n.d.). Historical electricity prices are taken from Nord Pool spot (Nord Pool Spot, n.d.) and are averaged between two price zones, namely DK1 (western Denmark) and DK2 (eastern Denmark). Data on solar production originate from the national aggregate provided by the Danish TSO Energinet (Energinet DK, n.d.). The performance is measured using the COP, which determines the ratio between the heat energy produced and the electricity input. This study uses a conservative approach by implementing observed efficiencies from (Poulsen et al., 2017). There, the COP varies according to the temperature profile.

As mentioned at the beginning of this section, the chosen year used a flat volumetric grid tariff and tax rates which allow a consumption profile to be observed unaffected by new varying tariffs. This study does not consider any flat annual subscription fees. The transmission grid tariff is levied by the transmission system operator Energinet and amounts to 8.3 Øre/kWh (0.011 €/kWh) (Energinet DK, 2017). The distribution system tariff is taken from an average Danish distribution system operator, returning 18.5 Øre/kWh (0.025€/kWh)(Trefor and El-net, 2017). Electricity for residential consumers is taxed at a volumetric rate of 91 Øre/kWh (0.12€/kWh) through the electricity taxation facility "Elafgift" (skat.dk, 2022). Taxes and grid tariffs are only imposed on net consumption from the grid.

*4.2 Investigated scenarios*

Four scenarios are optimized using the model presented in the previous subsection. Three scenarios cover the study's main results, and one describes a sensitivity analysis. Table 2 summarizes the investigated scenarios and naming conventions.

| Scenario name | BAU | NTAX | NTAX38 | NTAXHi |
|---|---|---|---|---|
| Scenario characteristics | Tax exemption for self-consumed electricity from PV production and double taxation | Uniform taxation of electricity consumption with equal tax level to BAU | Uniform taxation of electricity consumption with 38% reduced electricity tax level maintaining revenue neutrality for the | All consumption treated equally with equal tax level to BAU (and BAUHi) and electricity prices according to 2022 levels |

|  | state compared to BAU |
|--|--|

*Table 2. Overview of investigated scenarios*

The BAU scenario represents the current taxation scheme for Denmark, where self-consumed PV production is not subject to taxation, and battery arbitrage with the grid is taxed. The NTAX scenario taxes net imports and all PV production that is self-consumed or stored in the battery. Net imports going into the battery are also still taxed. However, when an equivalent amount of energy is sold to the grid, the household receives the paid taxes on top of the wholesale prices. The tax level equals the BAU scenario, thus leading to a higher total tax income as more flows are now taxed. The NTAX38 scenario reduces the tax rate by 38%, yielding a total tax income equal to the total of the BAU scenario once the broader tax base is considered. It is important to note that, since the number of households per category is equally sized in this study and not scaled according to the actual Danish household numbers, 38% does not correspond to the actual adjustment of the electricity tax in the Danish case. At the time of this study, the European electricity system is under pressure due to a gas supply shortage leading to a significant rise in electricity prices. To account for the higher general level of prices and volatility, the electricity prices are scaled up to match the 2022 level (between January and June) in a sensitivity analysis called NTAXHi. This sensitivity calculates the impact of higher prices on flexibility provision by the households to benefit the system.

## 5. Results

The four subsections in the results summarize the outcomes of the scenarios in the following order. Section 5.1 presents the total cost of electricity of all households in the BAU scenario, including the shares of taxation, grid tariffs and electricity costs, followed by the relative difference between NTAX and NTAX38 compared to BAU. Section 5.2 analyzes the total differences in the yearly cost and focuses on the redistribution across the socio-economic categories. Section 5.3 compares the households' electricity flows, detailing the equipment's operational changes. Finally, section 5.4 presents the sensitivity of the households in the BAU and NTAX scenarios with a higher price to validate the effects of equipment operation and the potentials of the new taxation scheme.

*5.1. Shifting the burden of electricity costs by taxing self-consumption*

The BAU scenario acts as a comparative base case in this study, which envisages a future with households as fully developed prosumers. Figure 5 shows all households sorted in ascending order by their yearly total cost of electricity, including the part that is paid in taxes (green), grid tariffs (orange) and the electricity cost (blue).

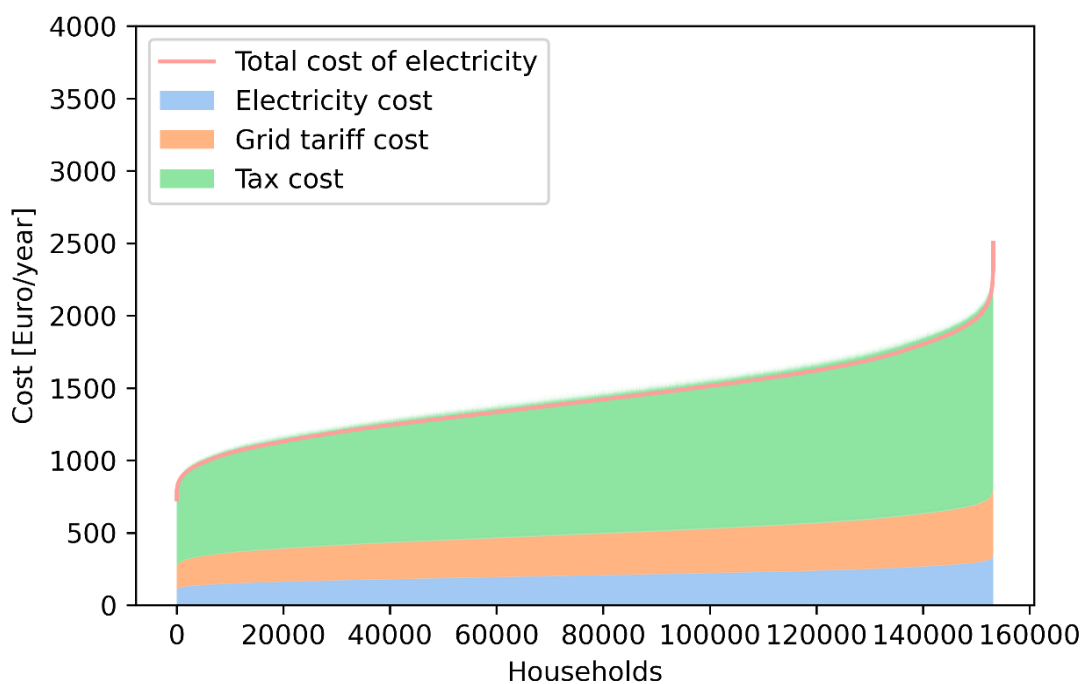

*Figure 5. Total cost of electricity for all houses sorted ascending.*

The yearly total cost of electricity ranges between €731 and €2502, while the largest share of the cost is due to taxes. The second largest cost component is grid tariffs, followed by the cost of electricity. Generally, the share of self-consumed electricity is between 30-44%. As self-consumption has a marginal cost of zero, it is not reflected in Figure 5. The distribution shows a relatively constant rise in cost between 10% and 90%, whereas the first and last thousand households have much lower or much higher costs.

Figure 6 and Figure 7 show the relative cost difference after implementing the uniform electricity taxation in NTAX and NTAX38 respectively. The order of the households remains the same as in Figure 5 to show that the households are affected differently even with similar total cost levels.

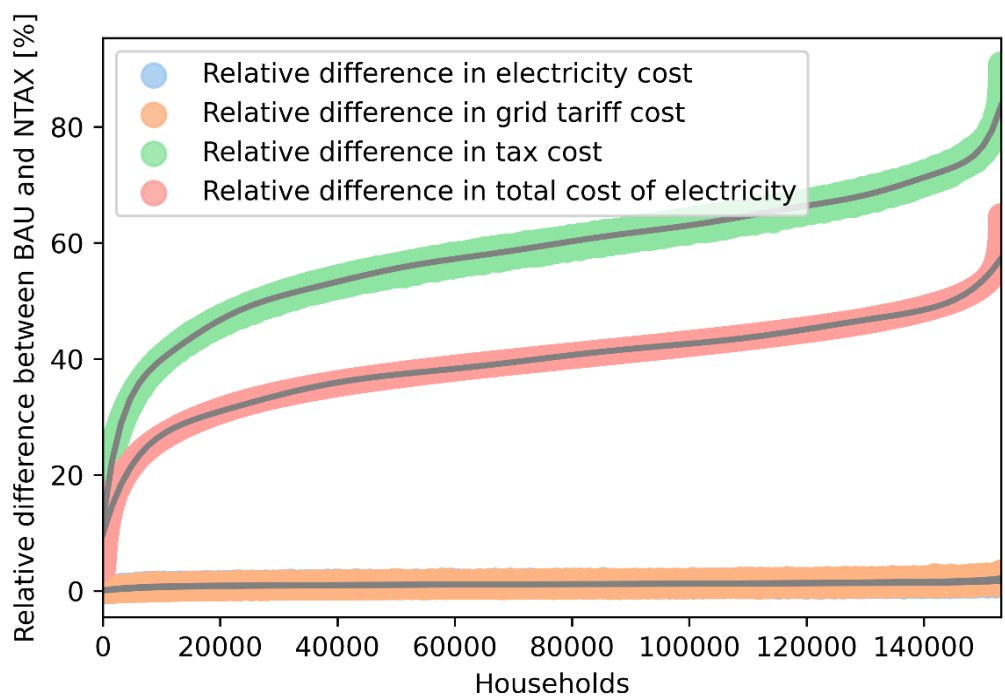

*Figure 6. Relative difference in cost and cost components between BAU and NTAX. The orange line hides the relative difference in electricity cost (blue line).*

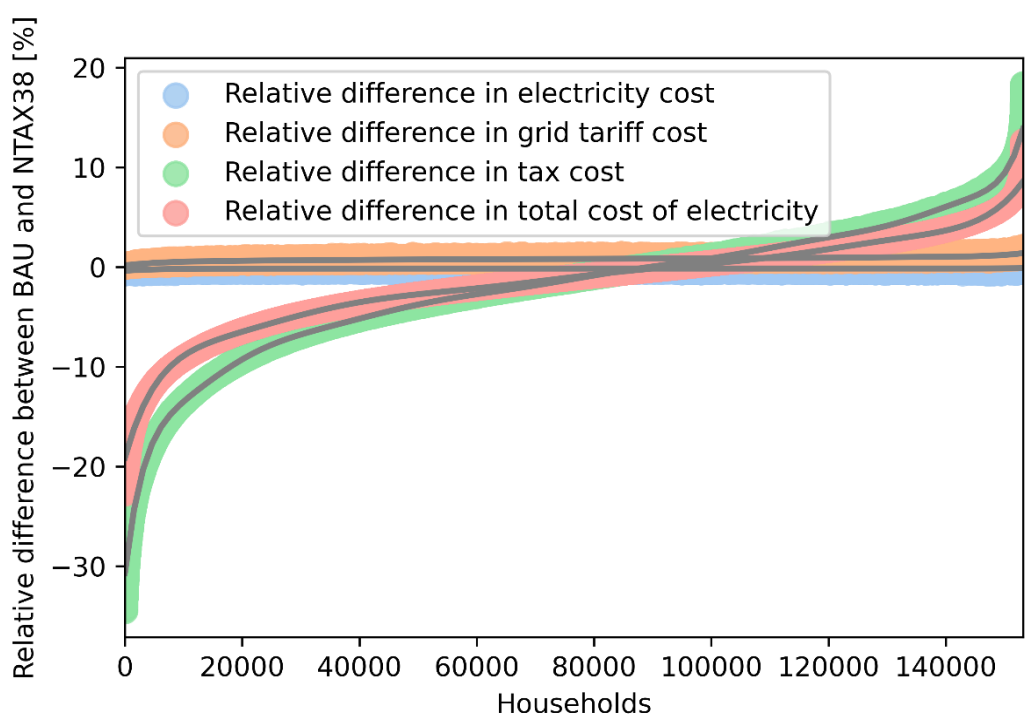

*Figure 7. Relative difference in cost and cost components between BAU and NTAX with a 38% reduction in the electricity tax*

Figure 6 shows that taxing all consumption raises tax expenses by 6-91% (green line), while the total cost also rises by 4-65% compared to the BAU scenario. The blurred lines show that the new taxation affects households differently along the line. While some households had almost identical yearly costs in the base case, they can now differ by up to 10% (adjacent in terms of the ordering of houses in Figure 5, e.g. house 10 versus house 11, where house 11 has a higher cost than house 10). This heterogeneity results from varying inputs regarding the timing and total amount of electricity consumption, as well as the EV availabilities.

In the NTAX38 scenario shown in Figure 7, which reduces taxes by 38% to obtain the same total annual revenue as BAU, the general cost increase for all households is no longer present. The share of the tax on the total electricity bill can now be reduced by up to 38%. Conversely, some households on the right-hand side can face increases of up to 18%. Similarly, the total cost of electricity per year also changes. Households with lower total bills can save up to 23%, while the households with the highest cost may face an increase of up to 13%.

Ultimately, uniform electricity taxation creates a level playing field between grid imports and self-consumed electricity, leading to a cost redistribution favoring smaller households. The change is mostly due to differences in taxation, as spending on grid tariffs and electricity costs from grid imports stay almost constant. Consumers with already large electricity bills but also owning a PV have to pay more compared to BAU when the tax level is reduced to match the total tax revenue. Conversely, around 61% of consumers reduce their yearly expenses for electricity. The next subsection describes which consumer groups face changes in their grid bills and which characteristics are essential for the differences.

*5.2. Socio-economic factors affecting yearly grid payments for consumers*

The NTAX and NTAX38 scenarios envisage a redistribution of cost. However, ordering households based on the total cost of electricity covers the internal dynamics and impacts that households' socio-economic characteristics have on the outcomes. Unraveling such dynamics is of the utmost importance to increase the transparency of policy changes and improve public acceptance of new initiatives. Figure 8 thus disaggregates the effect of NTAX depending on dwelling area and occupancy, and visualizes the average difference in the yearly total cost of electricity between the NTAX and BAU scenarios for twelve groups. Positive values correspond to cost increases originating from increasing the amount of taxable energy. Figure 13, Figure 14 and Figure 15 in the Appendix also visualize the average disaggregated based on the three income groups.

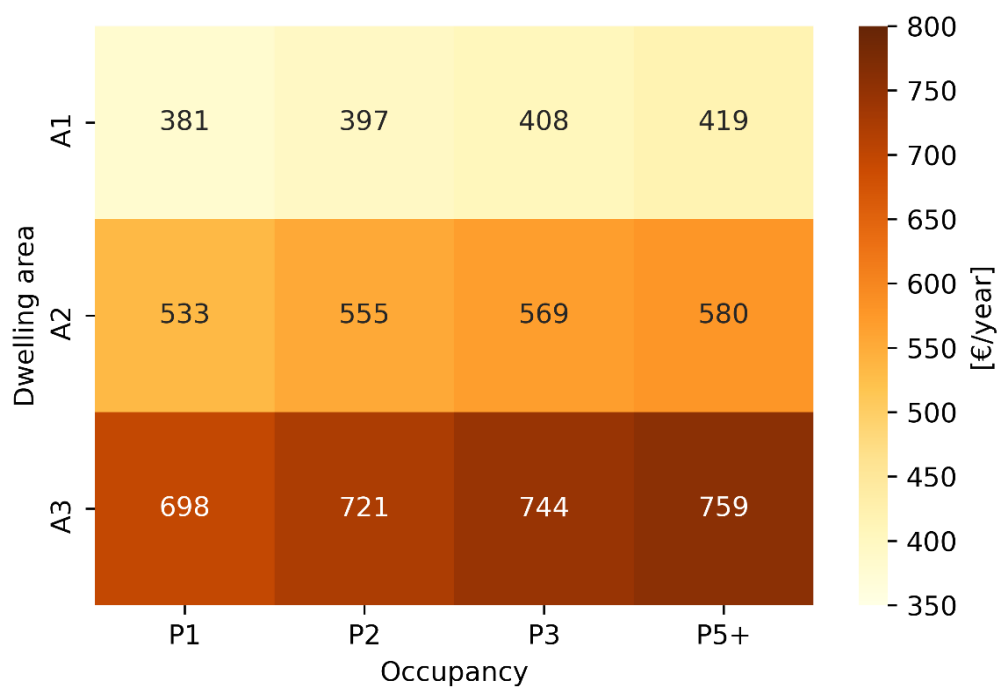

*Figure 8. Average difference of total yearly cost for electricity between the NTAX and BAU scenarios per socio-economic category. Positive values correspond to an increase in yearly cost with the NTAX scenario in €/year.*

Figure 8 shows that canceling the tax exemption for self-consumed solar PV production increases average yearly electricity bills by at least €381 /year to €759 /year, dependent on the group. The difference in yearly cost shows a clear trend: the larger the dwelling area and the number of occupants, the higher the cost increase. This pattern derives from the amount of self-consumed solar PV production. The larger this share, the higher the taxes that suddenly need to be paid after introducing the NTAX policy. When comparing single households to 5+ households in the same dwelling area category (and thus with a similar installed PV capacity), the quantity of self-consumed electricity from PV increases by around 10% for the latter group. Electricity consumption also rises generally with larger dwelling areas, regardless of the occupancy. Consequently, the share of self-consumed electricity peaks in the highest area and occupancy category. At the same time, it is notable that the effect of increased occupancy is marginal compared to the dwelling area. The component that changes the yearly electricity bill the most is the taxes, whereas the grid tariff and electricity cost expenses barely change (see Figure 7). While the difference in cost originating from the introduction of a self-consumption tax can be seen as a simple rise in the yearly bill, we would like to note that the outcomes can also be interpreted as the yearly avoided tax payments by households in the current BAU policy scheme. In the end, the NTAX scenario penalizes large consumers in particular. With the increased income generated from electricity taxes in the NTAX scenario, the state can reduce the general tax level by following revenue neutrality principles. Figure 9 therefore visualizes the cost redistribution of the NTAX38 scenario compared to BAU.

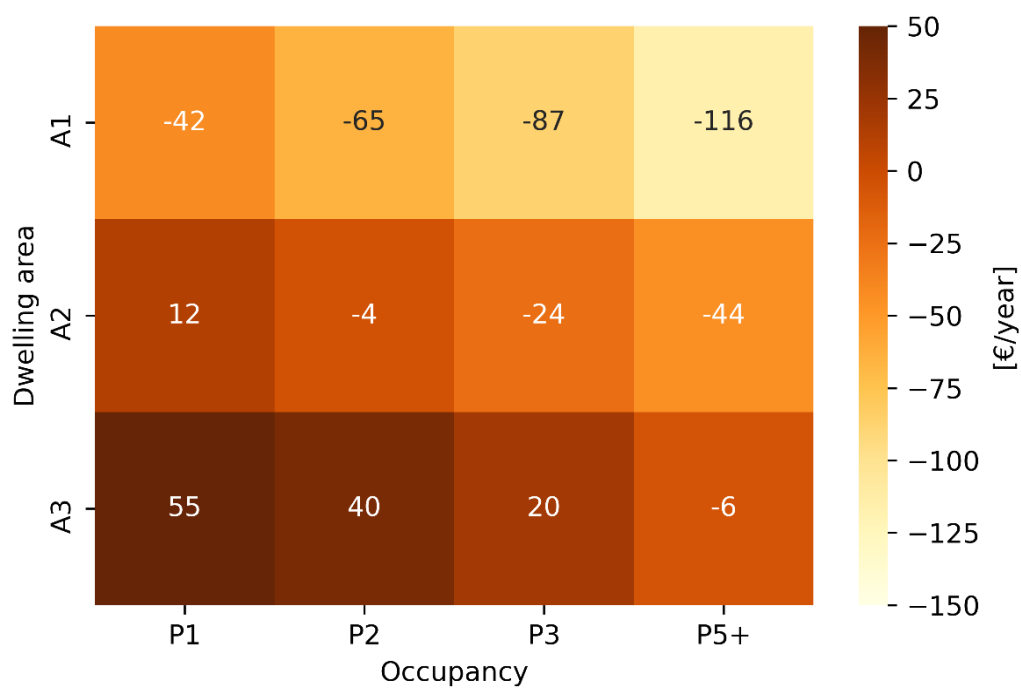

*Figure 9. Average difference of total yearly cost for electricity between the NTAX38 and BAU scenarios per socio-economic category. Positive values correspond to an increase in the yearly cost with the NTAX38 scenario in €/year.*

In contrast to NTAX, the NTAX38 scenario shows a different pattern. At first, Figure 9 shows that, while some household categories still face higher costs, most of them reduce their yearly expenses compared to the BAU scenario. Generally, the smaller the house and the higher the occupancy, the more important the cost savings of up to €116 /year. Meanwhile, singles living in medium and large houses and two persons living in large houses pay up to €55 /year more. The overall redistributive pattern of NTAX38 derives from the original dependence on the grid. Households with high rates of occupancy living in a small house with only small solar PV production require more direct imports from the grid to cover the demand. In contrast, households with low occupancy living in large houses are self-sufficient to a greater degree. The share of demand covered by PV for P1 in combination with A3 is approximately 43%, while P5+ living in an A1 house covers only 30% of the demand with self-produced electricity. Thus, introducing a tax on self-consumption while at the same time reducing the general tax level improves the relative economics of grid consumption. Consequently, households with a high occupant density can reduce their yearly electricity bills at the expense of those with a low occupant density with NTAX38.

While NTAX increases the cost for everyone, particularly for large consumers utilizing solar PV the most, reducing the tax level in NTAX38 decreases the cost for households the smaller the living area per person is. NTAX penalizes all consumers, whereas reducing the tax level in NTAX38 results in net benefits for several consumers at the expense of households with large living areas and low occupant densities.

The changes in the yearly grid bills come not only from the new taxation policy but also from adapting the optimal operation of the technologies. The next subsection therefore describes the effects of taxation on the utilization of technologies.

### 5.3. Effects of taxation on the utilization of technologies

Table 3 shows the averaged differences in electricity flows from supply to demand for the NTAX scenarios compared to BAU, whereas Table 4 shows the difference between NTAX38 and BAU. The changes in electricity flow present how components are used differently to visualize changes in flexible behavior and potential system efficiency gains originating from a new taxation policy.

| From \To | MP | PV | BT | CO | EV | HP | Sum |
|---|---|---|---|---|---|---|---|
| MP | 0 | 0 | 3.3 | 32.6 | 26.9 | 29.1 | 91.9 |
| PV | 134.9 | 0 | -113.3 | 0.3 | -4 | -18 | -0.1 |
| BT | -26.9 | 0 | 0 | -32.8 | -22.9 | -16.7 | -99.3 |
| Sum | 108 | 0 | -110 | 0 | 0 | -5.6 | -7.5 |

*Table 3. Average difference in flows of all households when switching from BAU to NTAX. Positive values correspond to an increased yearly flow in NTAX compared to BAU in [kWh/year].*

| From \ To | MP | PV | BT | CO | EV | HP | Sum |
|---|---|---|---|---|---|---|---|
| MP | 0 | 0 | 108 | -34.2 | 14.5 | -23.7 | 64.6 |
| PV | 84.8 | 0 | -63.4 | -3.4 | -6.4 | -11.8 | -0.2 |
| BT | -24.4 | 0 | 0 | 37.5 | -8 | 35 | 40.1 |
| Sum | 60.4 | 0 | 44.6 | 0 | 0 | -0.5 | 104.5 |

*Table 4. Average difference in flows of all households when switching from BAU to NTAX38. Positive values correspond to an increased yearly flow in NTAX38 compared to BAU in [kWh/year].*

Table 3 and Table 4 show flows from row to column as the difference from NTAX to BAU and NTAX38 to BAU respectively. Introducing taxation on self-consumption increases direct imports from the electricity grid by 91.9 kWh/year in NTAX and 64.6 kWh/year in NTAX38 respectively. At the same time, direct exports of solar PV increase by 13% and 9%. With the taxation of self-consumed electricity, energy imports gain competitiveness compared to domestically produced electricity to cover consumption needs. This dynamic also results in a greater feasibility of exporting electricity during hours with high prices. Households are consequently increasingly utilizing cheap grid imports while exporting more self-produced PV electricity when prices are high.

NTAX generally reduces usage of the stationary battery. The reason for that lies in the decrease in the competitiveness of behind-the-meter demand coverage. BAU mostly uses the battery to store self-produced electricity to avoid grid tariffs and taxes. Self-consumption taxation reduces the benefit of storing electricity behind the meter. However, the reduction of the tax level in NTAX38 shows another pattern, whereby battery use increases compared to the BAU scenario, as seen in Table 4. The origin of the increased battery usage is not be found in the increased charging from domestic solar PV production but rather results from higher grid imports utilizing lower prices in the electricity market. The battery in this case stores cheap electricity to cover the inflexible basic electricity demand and less flexible heat demand. NTAX38 demonstrates that the battery plays an active role in increasing grid flexibility to meet inflexible residential demand, whereas in BAU, there is a natural incentive for self-consumption rather than prioritizing system flexibility.

In the end, taxing the self-consumption of domestic solar PV production in NTAX increases direct interactions (more direct PV export, less battery usage), while a reduction in the tax level in NTAX38 increases flexibility compared to BAU. Taxing self-consumption and simultaneously reducing the tax level while maintaining overall tax income increases grid-side flexibility while reducing behind-the-meter consumption.

### 5.4. In the current context of higher prices

In current events of high electricity prices and variability, the role of flexibility is crucial to reduce high price fluctuations and the reliance on flexible but costly peak and backup capacities. Table 5 thus summarizes the effect of taxing self-consumption from domestic solar PV production on the utilization of technologies and flexibility for the NTAXHi scenario, which applies such high electricity prices.

| From\to kWh/year | MP | PV | BT | CO | EV | HP | Sum |
|---|---|---|---|---|---|---|---|
| MP | 0 | 0 | 667.3 | 382.9 | 251.2 | 558.4 | 1859.8 |
| PV | 1111.2 | 0 | -690.3 | -49.7 | -160.8 | -283.4 | -73 |
| BT | 667.8 | 0 | 0 | -333.1 | -90.3 | -264.9 | -20.5 |
| Sum | 1779 | 0 | -23 | 0.1 | 0.1 | 10.1 | 1766.3 |

*Table 5. Average difference in flows of all households when switching from BAUHi to NTAXHi both with high prices corresponding to the 2022 level in Denmark. Positive values correspond to an increased yearly flow in NTAXHi compared to BAUHi.*

The increased price fluctuations show in particular three main differences to Table 3. The battery charges significantly more from the grid, and the exports from the battery are increased by almost a factor of 10. Also, direct exports from the PV more than double. The high electricity prices mainly drive all changes in energy flows, while grid tariffs and taxes become less relevant. Consequently, households are more incentivized to interact with the system under the NTAXHi scenario. Stronger interactions with the grid can smooth out prices by increasing consumption when prices are low to cover demand and charge the battery. The battery can then be used for self-consumption or arbitrage. Exports of self-generated PV are more significant because the revenue from the market is higher than the savings from storing the electricity behind the meter.

## 6. Discussion

The discussion section is structured as follows. The first subsection compares and validates the presented results with similar studies from the literature and discusses the choice of methodology. The second subsection. puts the policy of taxing self-consumed electricity from solar PV into a spatial context with different bodies of legislation and derives the theoretical impact of the policy changes based on geographical factors and the current policy environment. The third subsection discusses taxing self-consumption while taking economic and technical efficiency, energy equity and the green transition into account.

### 6.1. Comparison of outcomes with the existing literature

According to the present results, the current electricity tax structure leads to cross-subsidies from households with lower annual electricity expenditure to those with a more expensive bill. This is a common finding in the literature, where prosumers reap private benefits at the expense of a larger cost burden for society as a whole, especially under volumetric grid tariffs that similarly apply to net imports (Boampong and Brown, 2020; Fett et al., 2019; Gautier et al., 2018; Manuel de Villena et al., 2021). (Ansarin et al., 2020) investigated different tariff structures for Austin (TX, USA), finding that flat volumetric tariffs, such as the one used for this analysis, lead to the largest cross-subsidies. While they investigated a transition to ToU tariffs or real-time pricing, our analysis shows that simply taxing self-consumption can also have the same effect. To further compare our results with a literature lacking studies on taxing self-consumption, we focus primarily on tariff studies and general prosumer optimization to validate our outcomes.

(Burger et al., 2020) model 100,000 individual households in Chicago under different tariff structures, finding that, while every tariff reform returns winners and losers, even a fixed tariff, irrespective of electricity consumption, leads to smaller distortions than volumetric tariffs. Along this axis, we also find that the expanded tax base, when taxing self-consumption, enables a reduction in the overall tax rate, resulting in most households benefitting, as their savings from the reduced tax rate exceed the additional expenditure to pay for the self-consumed electricity.

The shift towards taxing self-consumption can also address the current incentives towards increased self-consumption and reduced grid interactions to avoid taxes and tariffs (Heinisch et al., 2019; Strielkowski et al., 2017). While battery use is reduced under NTAX as imports increase, NTAX38 and NTAXHi both see increases in grid interactions. (Pena-Bello et al., 2017) can help explain the reduced battery use in the NTAX scenario, as they find that a flat volumetric tariff provides little incentive to use a battery. However, they also state that time-varying tariffs that encourage residential battery use mainly do so for purposes of price arbitrage, whereby exports to the grid are massively reduced.

Taxing self-consumption is shown to reduce cross-subsidies and encourage more flexibility. To similarly align household incentives with the system's interests more equitably and efficiently, other studies use a three-part tariff structure with a maximum demand charge levied on hours when the system is constrained (Boampong and Brown, 2020; Simshauser, 2016; Strielkowski et al., 2017).

Looking at the socio-economic dimensions, we observe the largest price increases from taxing self-consumption in houses with large surfaces and low occupancy. (De Groote et al., 2016) also find that a large dwelling area is linked with better suitability for solar PV under a system where self-consumption is not taxed. They further identify a correlation with higher occupancy, as this leads to higher electricity consumption and thus greater potential savings. However, this study only looks at the operation of solar PV and not investments, whereas theirs looks at the economic viability of solar PV based on household characteristics.

### 6.2. Critical review of the applied methodology and data

This study's methodological choice further impacts on outcomes and conclusions. As an exogenous input, demand is not adjusted to changes in electricity costs. Households with high tates of consumption especially could be expected to have some elastic demands, which would decrease under the NTAX scenario. This would reduce tax revenues but probably also increase exports. These changes are not expected to affect any conclusions regarding total cost and tax payment changes across consumer categories. Secondly, the methodological choice of using an operational optimization model means that the effect of the new tax policy on investments cannot be identified. Implementing uniform electricity taxes likely reduces the optimal size of investments in residential solar PV systems and batteries. Consequently, the difference in cost and savings in this study should be seen from the perspective of changing the policy scheme after prosumers have adopted the technologies in reality. The chosen sizes of technologies are further subject to sensitivity as opposed to investment optimizations, as acknowledged in the literature review and methodology section. At the same time, the choices are based on best practice reports from practitioners, ministries and the scientific literature to reflect the nature of household decisions being bound by budgetary constraints, being safety-driven and being as binary as possible (Esmat et al., 2022). The implementation of maximum capacities for households to be exempt from the tax on self-consumption further shows that a large incentive exists for households to oversize their solar PV system as long as the household budget allows them to do so. Therefore, an investment optimization model should also have a dynamic coupling to market prices depending on the collective residential solar PV adoption and on individual budget constraints to avoid always hitting the maximum allowed capacities on solar PV. Furthermore, the exogenous electricity prices do not reflect additional endogenous demand from heat pumps and electric vehicles. In reality, the market equilibrium and prices would be a result of a dynamic game between supply and demand. The random mixing of data on electricity, heat and electric vehicle demand could not take into account the possible co-linearities that might exist between demand sources affected by occupancy and individual behavior. Also, the discretization of the household categories (as described in Section 4.1.) averages out the progression of rises in heating and electricity consumption within categories by increasing dwelling areas.

In the context of an already changed policy of electricity taxes, the subsequent investments also impact on the total cost differences, avoided costs and total flexibility presented in this study. At the same time, taking the implementation of a flat tariff into account reduces the need for flexibility, limiting the impact on flexible behavior and the possibility of savings compared to passive consumers. The interplay of dynamic congestion signals and electricity prices may amplify the flexible behavior effects and further improve the technologies' operational utilization. Thus, the new taxation scheme has the potential to improve even further the required flexibility to reduce local congestion at its core rather than supporting defections from the grid.

*6.3 International relevance and generalization of the outcomes*

While the present case study delivers insights into the Danish case, the relevance and impact of the tax design are likely to be even more relevant in other countries. The proposed uniform tax design for all consumption is generally transferable, as the exemption from the taxation on domestically produced and consumed electricity is widely applied, as shown in the literature review and methodology section. As an exception, Germany has had a self-consumption surcharge since 2014. Still, it is at only 40% of the general surcharge under the Renewable Energy Act, and systems below 10 kWp can be exempt from this charge (Inderberg et al., 2018). In 2022 the maximum residential PV size was lifted to 30 kWp and therefore practically excludes all residential systems from charging self-consumption (BDEW Bundesverband der Energie- und Wasserwirtschaft e.V., 2022). (Fett et al., 2019) include this surcharge in their scenario analysis while trying to find the optimal PV and battery sizes from a prosumer's perspective. They only find a small effect from the inclusion of this surcharge, which is dwarfed by the feed-in tariff.

Still, policies can change rapidly, as the Spanish example shows. Spain had one of the most restrictive self-consumption regulations in the world after 2015. This policy framework was forcing households to feed-in surplus generation for free. At the same time, other sectors could obtain compensation for PV feed-in but had to pay backup charges for self-consumed energy, among other levies (López Prol and Steininger, 2017). Since a reform in 2019, all prosumers can sell their surplus electricity at wholesale prices, and self-consumption-specific charges have been removed. The same legislation also enshrined the "right to self-consume electrical energy without charges", thereby blocking any attempt at implementing a self-consumption charge (López Prol and Steininger, 2020). Consequently, especially in European energy systems and policy environments, the implicit subsidy on self-consumed electricity is in place at different sizes. A further discussion of the role of geography and the level of electricity taxation can be found in Appendix A.1.

The impact of the proposed uniform electricity tax is thus an interplay of two main factors. First, there is a potential degree of self-consumption and autonomy. Second, the height of the per-unit electricity tax affects the total of avoided taxes. The higher the applied electricity tax in combination with the potential self-consumed electricity from domestic production, the greater the impacts, assuming that prices are uniform and not distorted. Thus, this study is especially relevant in states with high electricity taxes (see Figure 2 in Section 2.2. for European examples) or where additional taxes and surcharges with volumetric tax characteristics apply.

### 6.4 Contribution to the discussion of economic efficiency, the green transition and energy equity

This study analyzes the potential future of prosumers producing and consuming behind-the-meter electricity under different electricity tax policy schemes. As a first priority, this study contributes to the discussion of creating a policy environment that ensures a least-cost, sustainable and equitable energy system for households by reviewing uniform electricity taxation. Uniform electricity taxes increase grid-side flexibility and efficiency and support smaller, occupant-dense households. These effects are aligned with current EU and US policy goals aiming to improve energy justice by supporting ordinary consumers.

We show that uniformly applied electricity taxes can reduce general tax levels. This change has positive effects on a majority of socio-economic household categories while especially helping households with a smaller footprint per inhabitant, whether because of a smaller total surface area or a larger occupancy. Around 3.5 times more low-income households live in small dwellings than in large dwellings in Denmark. In contrast, the majority of high-income households live in large dwellings in Denmark. Keeping the current tax regime would further increase the cross-subsidy from smaller and poorer households towards larger and well-off households. The outcomes of this study therefore suggest that implementing a uniform electricity tax design is economically efficient since it creates a level playing field and increases flexibility while at the same time improving energy equity. However, further studies are needed to prove that the results discussed for Denmark are also applicable other contexts.

Taxing self-produced electricity from domestic solar PV reduces the latter's feasibility. In light of the green transition, the taxation of self-consumed electricity and its advantages will delay the rapid adoption of fossil-free production technologies. Even though earlier studies have found that this impact might not be large (Tomasi, 2022), the negative economic signals should not be ignored. Avoided expenses are a key driver for the adoption of solar PV systems (Ali et al., 2021; Arnold et al., 2022b; Borenstein, 2022; Secretariat, 2020). However, the main barrier, particularly for vulnerable households, are budgetary constraints regarding the initial investment (European Union, 2022).

In the end, this discussion shows that uniform electricity taxes strike a balance between economic efficiency and energy equity versus incentivizing the adoption of green technologies. Taxing self-consumption implements a level playing field and improves fairness while reducing the feasibility of domestic solar production. When adopting green technologies is seen as very important, implementing a subsidy on investment costs could be recommended while acknowledging the possible disadvantages. With uniform electricity taxes, solar PV is most feasible in occupant-dense housing. Such households often have a restricted budget in Denmark and could thus draw an outsized benefit from taxing all consumption equally while subsidizing investments for solar PV (Gunkel et al., 2022).

As this study suggests, taxing self-consumed electricity, which in the public eye might seem like the long reach of the state into household economies, must be implemented with the highest possible degree of transparency and communication regarding its economic and social benefits. Also, shocks should be avoided, leading to a stepwise implementation pathway by increasing self-consumption taxes while decreasing general taxation over a certain period, such as a temporal dynamic policy adaption suggested by (Kitzing et al., 2020). Existing residential solar PV could continue to enjoy exemptions from electricity taxes while applying uniform taxation to newly installed PV systems. Furthermore, additional studies on the claims of budget constraints, subsidies and adoption dynamics are essential to see further if the suggested benefits of taxing self-consumption are valid.

**7. Conclusion**

This study has investigated the effect of applying electricity taxes to all residential consumption of prosumers, including the consumption met by rooftop solar PV production, while focusing on cost redistribution, efficiency and flexibility. An optimization model covers approximately 155,000 Danish households within 35 different socio-economic categories. Furthermore, the effects on consumers' electricity bills include the differences in the cost of electricity, electricity taxes and grid tariffs.

The investigation shows that uniform electricity taxation redistributes the costs from households with small electricity bills to those with larger ones. Since the broader tax base means a higher total revenue, the tax level in this case study can be reduced by 38% while keeping revenues stable. The subsequent reduction also progressively redistributes the cost, while 61% of all consumers see a reduction in their yearly bills.

When keeping current tax levels, single houses with a small dwelling area pay €381 /year more and households with a large dwelling area and high occupancy even up to €759 /year more. As the tax rate is reduced to keep revenues constant, the maximum increase is €55 /year for single occupants living in a large building, while occupant-dense households can save up to €116 /year. The redistribution thus improves the situation for occupant-dense housing at the expense of households with low occupancy rates and large living areas.

Implementing equal tax rates further changes the operation of the technologies in prosumer households. First, maintaining the tax rate and taxing self-consumption shifts the result away from behind-the-meter consumption and towards grid consumption. Total grid imports increase and less battery charging occurs, hinting that the battery's business case lies in autonomy in order to avoid taxes and grid tariffs. At the same time, direct exports rise as the incentive to store excess production disappears.

The tax reduction, in combination with uniform electricity taxes, results in a notable change in the operation of the technology. Households start to utilize the battery significantly more. While direct exports from solar PV still create revenue, the battery is charged more frequently using grid electricity. It covers especially less flexible demands such as base electricity consumption and the heat pump and utilizes more low-price hours in the electricity system. Taxing all consumption equally in combination with a tax reduction that maintains state revenue increases the interaction with the grid, possibly reducing system prices through higher exports and leveling low price hours with the battery imports. This scenario shows robust outcomes and even higher potential for the grid-side flexibility of households in the performed sensitivity analysis with the high European electricity prices of 2022.

The results of taxing all electricity consumption equally and disregarding the origin of its production show a clear tendency to make the tax burden fairer and increase flexibility and efficiency. Whether a uniform electricity tax is a good idea depends on how much people use solar power in a specific country, as well as how much they use electricity in general and how high the tax rate is. This means that the tax would affect different parts of Europe differently. Taxing all consumption diminishes the current regressiveness from electricity taxation and reduces the tax burden, particularly for vulnerable households.

This study shows that taxing behind-the-meter consumption, coupled with a revenue-neutralizing tax rate reduction, results in improved efficiency and a fairer tax burden. At the same time, the uniform taxation policy reduces the private economic feasibility of residential solar PV. Consequently, to assess the impact on the adoption rate of green technologies, future research should address the impact on PV adoption. Policymakers may address the reduced economic viability of solar PV once the tax exemption for self-consumption is removed and aligned with conditions for PV installations outside the household sector.

**Author contributions.**

Philipp Andreas Gunkel: conceptualization, methodology, software, analysis, writing

Febin Kachirayil: conceptualization, analysis, writing

Claire-Marie Bergaentzlé: conceptualization, writing, supervision

Russell McKenna: methodology, writing, supervision

Dogan Keles: methodology, writing, supervision

Henrik Klinge Jacobsen: methodology, writing, supervision

All authors have read and agreed to the published version of the manuscript.

**Funding.** This article was supported by the Flexible Energy Denmark (FED) project, funded by Innovation Fund Denmark (Grant No. 8090-00069B). In addition, it was partly supported by the FlexSUS project (No. 91352), which received funding in the framework of the joint programming initiative ERA-Net RegSus, with support from the European Union's Horizon 2020 research and innovation programme under grant agreement No. 775970.

**Conflicts of interest.** The authors declare no conflict of interest. The funders had no role in the design of the study, in the collection, analysis or interpretation of data, in the writing of the manuscript, or in the decision to publish the results.

**Appendix A**

*A.1. Additional characteristics of a defined subset of consumer categories*

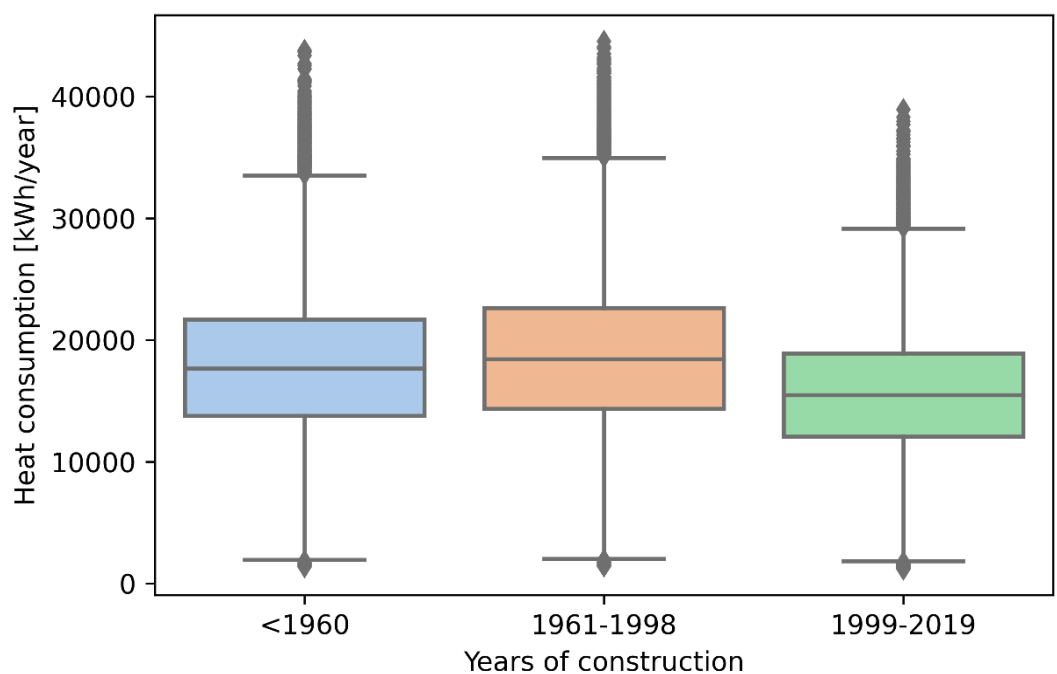

*Figure 10. Distribution of yearly heat consumption by households and construction year. The yearly heat consumption of Danish households is around 14000 kWh/year. The data shown do not indicate the scaled number of all Danish households, but rather represents the equal number of all (single family house) categories, thereby leading to an upwards shift.*

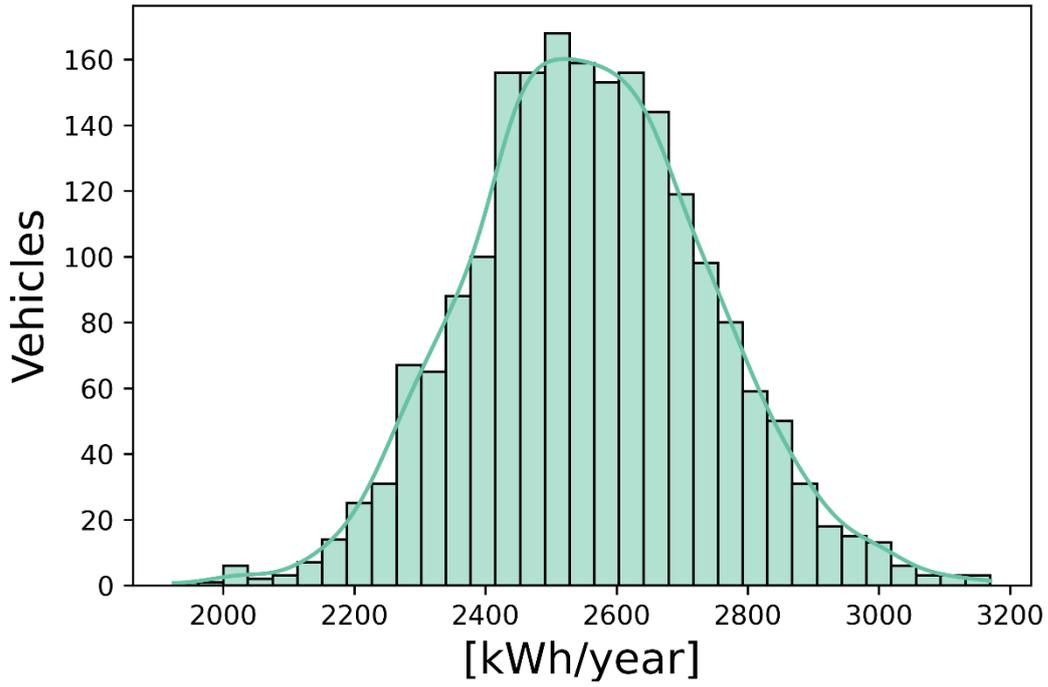

Figure 11. Yearly electricity consumption of the included EV fleet.

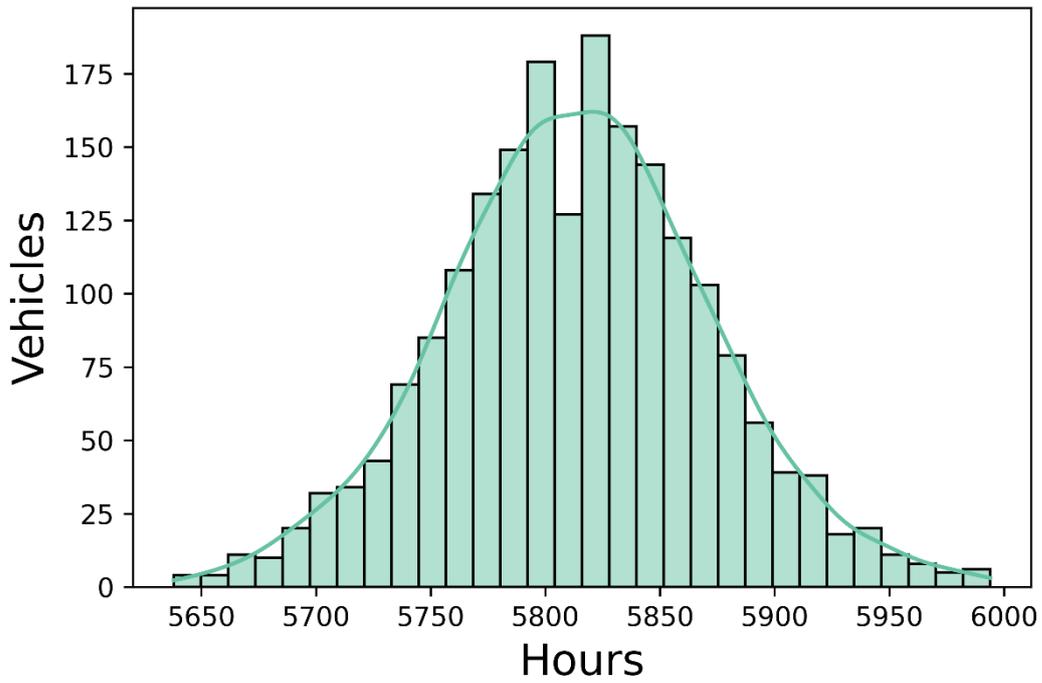

Figure 12. Yearly availability of the EV fleet. Availability is defined by being plugged into a residential charger.

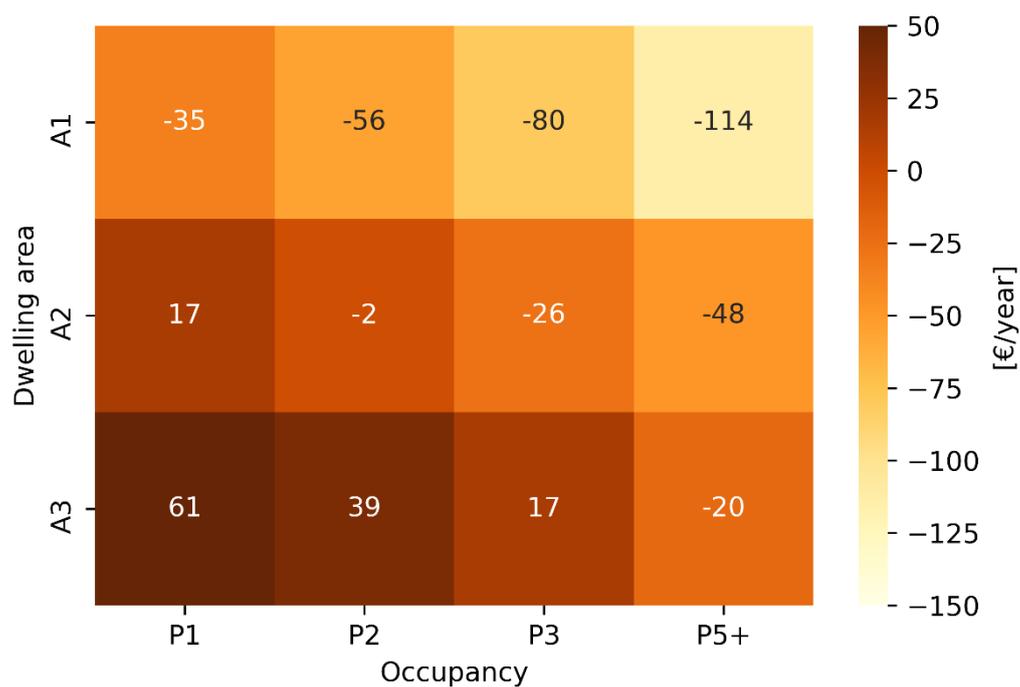

*Figure 13. Average difference in the annual cost of electricity of households in the low income group depending on the occupancy and living area category between the NTAX38 and BAU scenarios.*

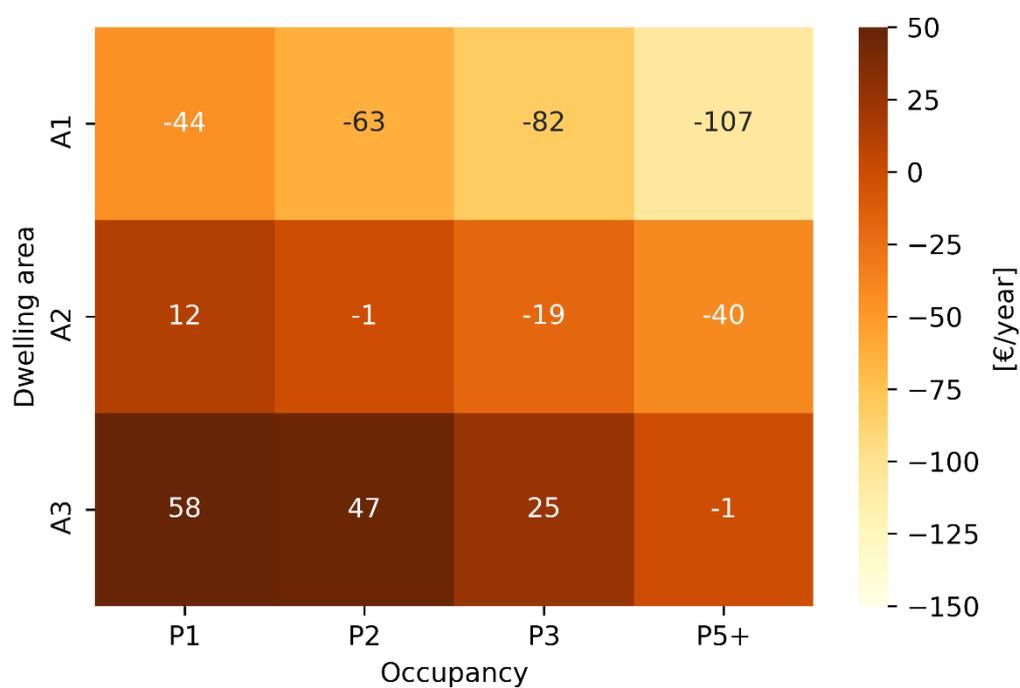

*Figure 14. Average difference in the annual cost of electricity of households in the middle income group, depending on occupancy and living area category between the NTAX38 and BAU scenarios.*

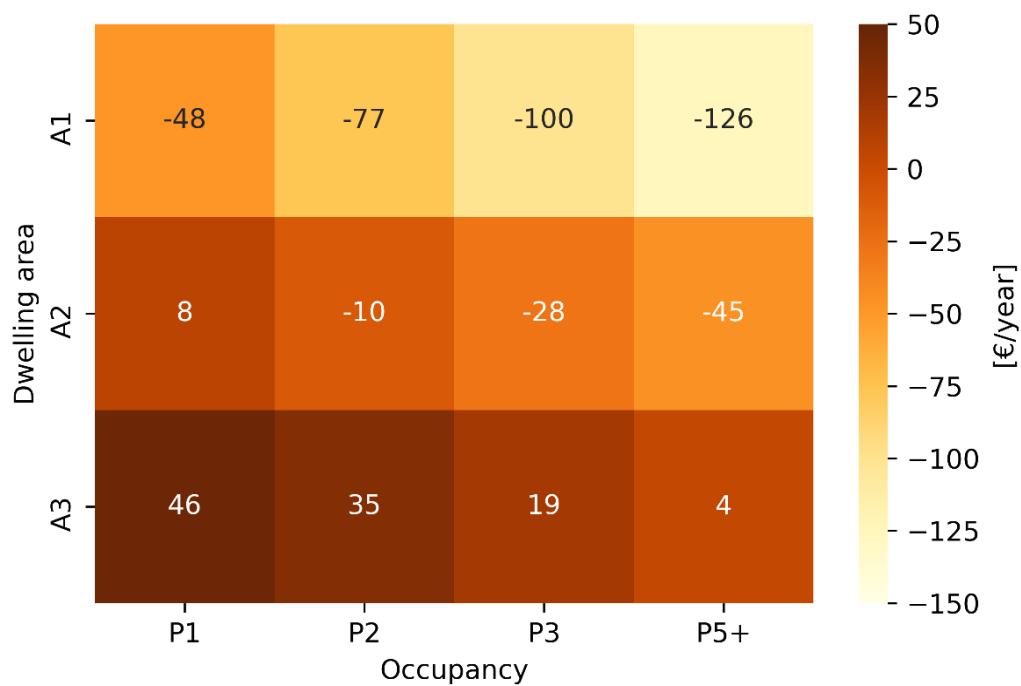

*Figure 15. Average difference in the annual cost of electricity of households in the high income group, depending on occupancy and living area category between the NTAX38 and BAU scenarios.*

**Discussion of the impacts from geography and the level of electricity taxation**

Irrespective of the individual policy environment, the positive impact of the proposed taxation scheme increases with the following technical factors and geographical scopes. At the core, the higher the solar potential in combination with a higher share of potential autonomy, depending on individual electricity profiles, the greater the avoided taxes and the incentives for behind-the-meter operations. The case study is located in Denmark, at a global latitude of around 55°. Denmark has a long-term average output of 1059 kWh/kWp per year with large seasonal differences (Solargis, 2022). Seasonal differences fall the more to the south the solar PV is located and the total potential output increases. Switzerland averages around 1241 kWh/kWp, whereas Spain achieves around 1534 kWh/kWp per year. Consequently, regarding the geographical potential of solar PV, the Danish case study lies at the far lower end. Furthermore, the fact that AC systems are not part of Danish residential consumption, in contrast to geographically more southerly countries, reduces the natural match between domestic solar production and electricity demand from AC.

In contrast, the electricity tax level is the second relevant determinant for regional differentiation and the subsequent generalization of the impact of taxing self-consumption. Danish electricity taxes are considerable compared to other countries, with around 12 ct/kWh (skat.dk, 2022). Switzerland has electricity taxes of around 4.2 ct/kWh (Swissgrid, n.d.), and the US does not apply specific federal taxes on residential electricity end-use besides the sales tax (U.S. Department of Energy, 2021). On the regional or county level, additional charges may apply. Examples of this are the Utility Usage Taxes or California Credit systems in the US, with characteristics comparable to taxation subject to the investigations in this study, while levels are still comparably low (Southern California Edison, n.d.). The lower the general tax levels applied to electricity consumption, the higher the shares of avoided taxes, and the fewer the flexibility barriers, which are less impactful.